\documentclass[twoside]{article}    

\usepackage{amsfonts,amssymb, amsthm, amsmath}
\newcommand{\tr}{\mbox{tr}}

\def\minipar{\par\vskip3pt\par}

\numberwithin{equation}{section}
\theoremstyle{plain}

\begin{document}

\pagenumbering{arabic}
\pagestyle{myheadings}
\markboth{\noindent\centerline{COECKE,
MOORE AND WILCE}}{\noindent\centerline{OPERATIONAL QUANTUM LOGIC: AN
OVERVIEW}}
\thispagestyle{plain}
\setcounter{page}{1}
\hbox{}
\par\vskip 2.25 truecm\par
\noindent{\bf OPERATIONAL QUANTUM LOGIC: AN OVERVIEW\footnote{This
paper is a slightly modified version of the introductory chapter to the
volume B.~Coecke, D.J.~Moore and A.~Wilce (Eds.), {\it Current Research
in Operational Quantum Logic: Algebras, Categories, Languages},
Fundamental Theories of Physics series, Kluwer Academic Publishers,
2000.  Consequently, there is a particular focus in this paper on
the (survey) articles of that volume, which are dedicated
to D.J.~Foulis in honor of his seminal contributions to our field.}}
\par\vskip 0.406 truecm\par
\par\vskip 0.406 truecm
\parindent=2cm\par
BOB COECKE\footnote{The author is Postdoctoral Researcher at Flanders'
Fund for Scientific Reseach.}
\par {\it Department of Mathematics,
\par Free University of Brussels,
\par Pleinlaan 2, B-1050 Brussels, Belgium.}
\par e-mail: bocoecke@vub.ac.be
\par\vskip 0.406 truecm\par
DAVID MOORE\footnote{Current address: Department of Physics and
Astronomy, University of Canterbury, PO Box 4800 Christchurch, New
Zealand.}
\par {\it Department of Theoretical Physics,
\par University of Geneva,
\par 24 quai Ernest-Ansermet, CH-1211 Geneva 4, Switzerland.}
\par e-mail: d.moore@phys.canterbury.ac.nz
\par\vskip 0.406 truecm\par
ALEXANDER WILCE
\par {\it Department of Mathematics and Computer Science,
\par Juniata College,
\par Huntingdon, PA 16652, USA.}
\par e-mail: wilce@juniata.edu
\par\vskip 0.406 truecm\par
\par\vskip 0.406 truecm\par
\par\vskip 0.406 truecm\par

\parindent=0.6cm

\noindent The term quantum logic has different connotations for
different people, having been considered as everything from a
metaphysical attack on classical reasoning to an exercise in
abstract algebra. Our aim here is to give a uniform presentation of
what we call operational quantum logic, highlighting both its concrete
physical  origins and its purely mathematical structure. To orient
readers new to  this subject, we shall recount some of the historical
development of quantum  logic, attempting to show how the physical and
mathematical sides of  the subject have influenced and enriched
one another.

\vfill\eject


\noindent  {\bf 1. Introduction}
\setcounter{section}{1}
\setcounter{thm}{0}
\setcounter{equation}{0}
\par\vskip 0.406 truecm\par
\par\noindent
The subject of {\sl operational quantum logic} --- lying somewhere at
the crossroads of mathematics, physics and philosophy --- has a long
and complicated history, and has generated a large, scattered and
unruly literature. It is not an easy thing to explain, in a few words,
what it is supposed to be about! Our best attempt is to say that
operational quantum logic involves
\begin{itemize}
\item[(a)] the fact that the structure of the 2-valued observables in
orthodox quantum mechanics may usefully be regarded as a
non-classical propositional logic,
\item[(b)] the attempt to give independent motivation for this structure, as
part of a general programme to understand quantum mechanics, and
\item[(c)] the branch of pure mathematics that has grown out of (a) and (b),
and now concerns itself with a variety of ``orthomodular''
structures generalizing the logic of 2-valued quantum observables.
\end{itemize}

\minipar

Whatever else it may be, quantum logic is a living, growing part
of contemporary mathematics and theoretical physics ---  one that
has continued to hold the interest of a body of mathematicians,
physicists and philosophers of science. This sustained interest
reflects in part the fact that the basic ideas and language of
quantum logic inform most discussions of the vexing foundational
problems of quantum mechanics (indeed, to a degree that is often
not recognized even by the discussants). It reflects also the fact
that quantum logic has spawned an autonomous and fascinating
branch of pure mathematics concerned with a variety of structures
--- orthomodular lattices and posets, orthoalgebras, partial
Boolean algebras, etc. --- generalizing
${\Bbb P}({\bf H})$. Finally, the recent advent of quantum
computation and quantum information theory offers a field of
practical applications for quantum logic which has yet to be fully
explored.

\minipar

Historically, quantum logic derives from von Neumann's (more than
casual) observation that the 2-valued observables, represented in
his formulation of quantum mechanics by projection operators,
constitute a sort of  ``logic'' of experimental propositions. This
idea was further pursued by Birkhoff and von Neumann. After two
decades of neglect, interest in quantum logic was revived, due in
large part to Mackey's analysis of the probabilistic calculus of
standard quantum theory coupled with his theory of induced
representations. The further development of the subject has
occurred at several levels and in a number of directions. Mackey's
work was extended significantly by Piron, whose representation
theorem and axiomatic framework provided much impetus for further
development. At the same time, dissatisfaction with Mackey's
axiomatic framework led to a search for more primitive, and more
concretely operational, foundations. Notable here is the work of
Foulis and Randall, and also that of Ludwig and his colleagues at
Marburg. Work in foundational physics has also stimulated, and
been stimulated by, purely mathematical research, notably in the
development of an abstract theory of orthomodular lattices and, in
recent years, more general structures such as orthoalgebras and
effect algebras. More recently still, the subject has seen the
application of powerful category-theoretic techniques. Given the
somewhat overwhelming variety of these developments, in this
introductory essay, we are going to attempt an outline of quantum
logic that will help readers who are not already experts in the
subject to understand the various papers that are produced --- and
also, to see them as belonging to a common subject. We start by
discussing the seminal work of Birkhoff and von Neumann and its
development by Mackey. We then turn to a brief exposition of
Piron's representation theorem and his axiomatic framework,
referring the reader to
[Coecke and Moore 2000] and [Valckenborgh 2000] for
expositions in a more contemporary, categorical idiom. Next, we discuss
the work of Foulis and Randall,
in particular their introduction of the notion of an orthoalgebra,
and their observation that tensor products of
orthomodular posets generally exist only as orthoalgebras. The
Foulis-Randall formalism is discussed in
detail in [Wilce 2000].
We follow with a general exposition of the pure
mathematical theory of orthomodular structures. For more details on
orthomodular lattices we refer to [Bruns and Harding 2000], for
observables on orthomodular posets to [Ptak 2000] and for group
representations  on (interval) effect algebras to [Foulis 2000].
Finally, we mention the notion of categorical enrichment and the theory
of quantales, surveyed respectively in [Borceux and Stubbe 2000] and
[Paseka and Rosick\'y 2000], before introducing computational and
linguistic aspects, treated respectively in [Resende 2000] and
[Gudder 2000].

\vskip 0.406 truecm\par
\par\vskip 0.406 truecm\par
\noindent
{\bf 2. Von Neumann's quantum mechanics}
 \setcounter{section}{2}
\setcounter{thm}{0}
\setcounter{equation}{0}
\par\vskip 0.406 truecm\par
\par\noindent
While precise mathematical treatments of quantum mechanics existed
before Johann von Neumann's monumental treatise [1932], it may
reasonably be argued that this work fixed once and for all the
theoretical framework of standard quantum theory, in which each
quantum mechanical system is associated with a Hilbert space ${\bf
H}$, each unit vector $\psi \in {\bf H}$ determines a {\sl state}
of the system, and each observable physical quantity associated
with the system is represented by a self-adjoint operator $A$ on
${\bf H}$. The spectral theorem tells us that such an operator is
associated with a {\sl spectral measure}
$$P_{A} : {\cal B}({\Bbb R}) \rightarrow {\Bbb P}({\bf H})$$
assigning to each real Borel set $B$ a projection operator
$P_{A}(B)$ on ${\bf H}$. For any unit vector $\psi \in {\bf H}$, the
quantity
$$\mu_{A,\psi}(B) := \langle P_{A}(B)\psi, \psi \rangle$$
then defines a probability measure on the line, which von Neumann
regards as giving the probability that the observable (represented
by) $A$ has a value in the set $B$ when the system's state is
(represented by) $\psi$. If the identity function has finite
variance in $\mu_{A,\psi}$, then $\psi$ is in the domain of $A$,
and the expectation value of $A$ relative to $\psi$ is given by
$$\mbox{Exp}(A,\psi) = {\textstyle\int_{\Bbb R}}\,s\,d\mu_{A,\psi}(s).$$
One easily checks that this works out to
$\mbox{Exp}(A,\psi) = \langle A \psi, \psi \rangle\,.$
\par\vskip 0.456 truecm\par
\par\vskip 0.406 truecm\par
\noindent
{2.1. THE LOGIC OF PROJECTIONS}
\par\vskip 0.406 truecm\par\noindent
If the {\sl Mathematische Grundlagen der Quantenmechanik\/}
signalled the passage into maturity of quantum mechanics, it also
signalled the birth of quantum logic. Evidently, it is the
projection-valued measure $P_{A}$, more than the operator $A$,
that most directly carries the statistical interpretation of
quantum mechanics outlined above. Now, as von Neumann notes, each
projection $P \in {\Bbb P}({\bf H})$ itself defines an observable
--- one with values $0$ and $1$. If $P = P_{A}(B)$ is the spectral
projection associated with an observable $A$ and a Borel set $B$,
we may construe this observable as ``testing'' whether or not $A$
takes a value in $B$. Von Neumann regards $P$ as representing a
physical property of the system (or rather, of the system's
states). He remarks that
\begin{quote}
``the relation between the properties of a physical system on the
one hand, and the projections on the other, makes possible a sort
of logical calculus with these. However, in contrast to the
concepts of ordinary logic, this system is extended by the concept
of `simultaneous decidability' which is characteristic for quantum
mechanics.'' [von Neumann 1932, p.253]
\end{quote}
Indeed, if $P$ and $Q$ are commuting projections, then their meet
$P\wedge Q$ and join $P \vee Q$ in the lattice ${\Bbb P}({\bf H})$
may be interpreted classically as representing the conjunction and
disjunction of the properties encoded by $P$ and $Q$; further, the
projection $P' = 1 - P$ serves as a sort of negation for $P$. If
$P$ and $Q$ do not commute, however, then they are not
``simultaneously decidable'', and the meaning of $P \wedge Q$ and
$P \vee Q$ is less clear. Nevertheless, ${\Bbb
P}({\bf H})$ retains many features of a Boolean algebra, and may
be regarded as an algebraic model for a non-classical
propositional logic.
In particular, ${\Bbb P}({\bf H})$ is orthocomplemented and so enjoys
analogues of the de Morgan laws; more sharply, the
sub-ortholattice generated by any commuting family of projections
is a Boolean algebra.
\par\vskip 0.406 truecm\par
\par\vskip 0.406 truecm\par
\noindent
{2.2. THE LOGIC OF QUANTUM MECHANICS}
\par\vskip 0.406 truecm\par\noindent
It is noteworthy that von Neumann speaks of the  simultaneous
``decidability'' (i.e., testability) of properties, but does not
distinguish between decidable and undecidable properties {\sl per
se}. Classically, of course, {\sl any} subset of the state-space
counts as a {\sl categorical} property of the system, and nothing
in principle prevents us from taking the same view in quantum
mechanics. However, only those subsets of the state space
corresponding to closed linear subspaces of the Hilbert space are
associated with observables, and so ``decidable'' by measurement.
If one adopts a rather severe positivism, according to which no
{\sl un}-decidable proposition is meaningful at all, one is led to
the seemingly strange doctrine that, for a quantum mechanical
system, the set of meaningful properties forms, not a Boolean
algebra, but rather the lattice ${\Bbb P}({\bf H})$ of projections
of a Hilbert space.

\minipar

This idea was further developed by von Neumann in a joint paper
with Garrett Birkhoff entitled {\sl The Logic of Quantum
Mechanics} [Birkhoff and von Neumann 1936]. Birkhoff and von
Neumann observe that ${\Bbb P}({\bf H})$ retains a number of the
familiar features of the algebra of classical propositional logic
--- in particular, it is orthocomplemented and hence satisfies de
Morgan's laws. It is not, however, Boolean --- that is, the
distributive law fails. Birkhoff and von Neumann go so far as to
suggest that
\begin{quote}
``whereas logicians have usually assumed that properties L71-L73 of negation were
the ones least able to withstand a critical analysis, the study of
mechanics point to the {\it distributive identities\/} L6 as the weakest
link in the algebra of logic.'' [Birkhoff and von Neumann 1936, p.839]
\end{quote}
As we shall see in section 7, this remark is rather deeper than
one may imagine, being interpretable in terms of the fundamental
difference between Heyting algebras and orthomodular lattices
considered as generalizations of Boolean algebra.

\minipar

This suggestion that the projection lattice may be viewed as a
propositional logic has been understood in a number of rather
different ways. Some have seen it as calling into question the
{\sl correctness} of classical logic. Others  have seen it as
entailing a less drastic modification of classical probability
theory. As we have seen, von Neumann himself [1932 \S3.5] is
rather cautious, remarking that the equivalence between subspaces
and projections induces a {\sl sort} of logical calculus.
Similarly, Birkhoff and von Neumann [1936 \S0] conclude that by
heuristic arguments one can reasonably expect to find a calculus
of propositions for quantum mechanical systems which is {\sl
formally} indistinguishable from the calculus of subspaces and
{\sl resembles} the usual logical calculus.

\minipar

More radical is the view of Finkelstein [1968, 1972] that logic is
in a certain sense empirical, a view championed by such
philosophical luminaries as Putnam [1968, 1976]. Finkelstein
highlighted the abstractions we make in passing from mechanics to
geometry to logic, and suggested that the dynamical processes of
fracture and flow already observed at the first two levels should
also arise at the third. Putnam, on the other hand, argued that
the metaphysical pathologies of superposition and complementarity
are nothing more than artifacts of logical contra\-dictions
generated by an indiscriminate use of the distributive law.

\minipar

This view of the matter, which remains popular in some
quarters,\footnote{For instance, Bamberg and Sternberg [1990,
pp.833-835] write: ``In fact, [quantum mechanics] represents the
most profound revolution in the history of science, because it
modifies the elementary rules of logic. $\ldots$ [T]he
distributive law does not hold in quantum logic. As we mentioned
above, the validity of quantum mechanics has been experimentally
demonstrated over and over again during the past sixty years. So
experiment has shown that one must abandon one of the most
cherished principles of logic when dealing with quantum
observables.''} depends on a reading of the projection
$P$ as encoding a {\sl physical property} of the quantal system,
and on the assumption that only physical properties are ultimately
to count as meaningful (or at any rate, as fundamental). There is,
however, a different way of construing $P$, namely, that it
encodes a {\sl statement} about the possible result of some
``measurement''. Thus, if $A$ is the self-adjoint operator
corresponding to the observable ${\cal A}$ and $P = P_{A}(B)$ is the
spectral projection of $A$ corresponding to the Borel set $B$, we
might construe $P$ as encoding the proposition that a measurement
of ${\cal A}$ {\sl would} yield a value in $E$ {\sl if made}. This
construal, usually dubbed {\sl operational}, guided Mackey in his
reconstruction of von Neumann's quantum mechanics, to which we now
turn.

\vskip 0.406 truecm\par
\par\vskip 0.406 truecm\par
\noindent
{\bf 3. Mackey's programme}
\setcounter{section}{3}
\setcounter{thm}{0}
\setcounter{equation}{0}
\par\vskip 0.406 truecm\par
\par\noindent
In an influential paper [1957], subsequently expanded into the
monograph [1963], George Mackey argued (a) that one could
reconstruct most if not all of the apparatus of von Neumann's
quantum mechanics from the premise that the experimental
propositions form an ortholattice isomorphic to ${\Bbb P}({\bf
H})$, and (b) that this premise itself could be independently
motivated by very general considerations about how probabilistic
models of physical systems ought to look.
\par\vskip 0.406 truecm\par
\par\vskip 0.406 truecm\par
\noindent
{3.1. QUANTUM MECHANICS AS A PROBABILITY CALCULUS}
\par\vskip 0.406 truecm\par\noindent
Mackey construed quantum mechanics as simply {\sl being} a
non-classical probability calculus, in which the Boolean algebra
of events of classical probability theory is replaced by the
lattice ${\Bbb P}({\bf H})$. More exactly, Mackey stressed that
both the states and the observables of a quantum mechanical system
can be defined purely in terms of ${\Bbb P}({\bf H})$. First, any
statistical state $W$ determines a probability measure on
${\Bbb P}({\bf H})$, namely the mapping
$$\omega_{W} : {\Bbb P}({\bf H}) \rightarrow [0,1];\ P \mapsto \tr(PW).$$
A deep theorem by Gleason [Gleason 1957; Dvure\v{c}enskij 1993] shows
that, conversely, every
$\sigma$-additive probability measure on ${\Bbb P}({\bf H})$ has
this form. Second, an observable with values in the measurable
space $(S,{\frak F})$ may be represented by a projection-valued
measure
$$M : {\frak F} \rightarrow {\Bbb P}({\bf H})$$
where, for
each measurable set $B \in {\frak F}$, the projection
$M(B)$ is taken to encode the ``experimental proposition'' that a
measurement of the observable yields a value in the set $B$.
Evidently, we may pull probability measures on ${\Bbb P}({\bf H})$
back along $M$ to obtain a classical probability measure on
${\frak F}$. We interpret $M^{*}(\omega) = \omega
\circ M$ as giving the statistical distribution of values of $M$ (in $S$)
when the system is in the state represented by $\omega$. In other words,
$$\omega_{W}(M(B)) = \tr(M(B)W)$$
represents the probability that the
observable represented by $M$ will yield a value in the set $B$, {\sl when
measured}, when the state of the system is represented by $W$.

\minipar

This connects with von Neumann's operator-theoretic representation
of observables in a natural manner, as follows: if $f : S
\rightarrow {\Bbb R}$ is any bounded classical real-valued random
variable defined on $S$, we may define the self-adjoint operator
$$A_{f} := {\textstyle\int_{S}}\,f(s)\,dM(s)$$
in the usual
way.\footnote{If $f$ is non-negative, then $A_{f}$ is given by the
supremum of the operators $A_{g} = \sum_{i} g_{i} M(B_{i})$ where
$g = \sum_{i} g_{i}\chi_{B_{i}}$ is a simple random variable with
$0 \leq g \leq f$.} Then, for any probability measure $\mu$ on
${\Bbb P}({\bf H})$, we have
$$E_{M^{*}(\mu)}(f) = {\textstyle\int_{S}}\,f(s)\,dM^{*}(\mu)(s) =
\tr(A_{f}W)$$
where $W$ is the density operator corresponding to $\mu$. This
view of quantum mechanics is strikingly powerful. Gleason's
theorem, together with the spectral theorem, the classical results
of Stone, Wigner, Weyl and von Neumann, and Mackey's own work on
induced unitary representations, allow one essentially to {\sl
derive} the entire apparatus of non-relativistic quantum mechanics
(including its unitary dynamics, the CCRs, etc.), from the premise
that the logic of experimental propositions is represented by the
projection lattice ${\Bbb P}({\bf H})$. For an outline of this
reconstruction, see [Mackey 1963] or
[Beltrametti and Cassinelli 1981]; for a detailed account, see
[Varadarajan 1968].
\par\vskip 0.406 truecm\par
\par\vskip 0.406 truecm\par
\noindent
{3.2. MACKEY'S AXIOMS}
\par\vskip 0.406 truecm\par\noindent
Its success notwithstanding, Mackey's account of quantum mechanics
as a probability calculus still rests on one undeniably {\sl ad
hoc} element: the Hilbert space ${\bf H}$ itself.  Indeed, once
one entertains the idea that the testable propositions associated
with a physical system need not form a Boolean algebra, the door
is opened to a huge range of other possibilities. It then becomes
a matter of urgency to understand {\sl why} nature (or we) should
choose to model physical systems in terms of projection lattices
of Hilbert spaces, rather than anything more general. Mackey
outlined an ambitious programme to do just this, by deducing the
Hilbert space model from a set of more primitive and, ideally,
more transparently plausible axioms for a calculus of events.

\minipar

The framework Mackey adopts is an abstract structure $({\cal O},
{\cal S}, p)$, where ${\cal O}$ is understood to represent the set
of real-valued ``observables'' and ${\cal S}$ the set of
``states'' of a physical system. These are connected by a mapping
$$p : {\cal O} \times {\cal S} \rightarrow \Delta: (A,s) \mapsto
p_{A}(~\cdot~|s)\,,$$
where $\Delta$ is the set of Borel probability measures on the
line. The intended interpretation is that
$p_{A}(~\cdot~|s)$ gives the statistical distribution of values of
a measurement of the observable $A \in {\cal O}$, when the system
is in the state $s \in {\cal S}$. We may take the pair $(A,B)$,
where $A \in {\cal O}$ and $B$ is a real Borel set, to represent
the ``experimental proposition'' that a measurement of $A$ yields
(would yield, has yielded) a value in $B$. Mackey considers two
such propositions equivalent iff they have the same probability in
every state --- in other words, $(A_{1},B_{1})$ and
$(A_{2},B_{2})$ are equivalent iff the associated mappings
$P_{A_i,B_i} := p_{A_i}(B_i|~\cdot~)$ are the same. The set $L$ of
such mappings $P_{A,B}$, which he calls {\sl questions}, is
Mackey's quantum logic.

\minipar

Now, ordered pointwise on ${\cal S}$, the set $L$ is an
orthocomplemented poset with unit $1$ given by $P_{A,{\Bbb R}}$
for any observable $A$, whose orthocomplementation is given by
$P_{A,B}' = 1 - P_{A,B} = P_{A,{\Bbb R} \setminus B}$.
Let us say that the questions $P, Q \in L$ are {\sl compatible}
iff $P = P_{A,B}$ and $Q = P_{A,C}$ for some common observable $A$
and some pair of Borel sets $B$ and $C$. Then we may consider $P$
and $Q$ to be  ``simultaneously measurable''. Further, let us say
that the questions $P, Q \in L$ are {\sl orthogonal} (or
``disjoint'', in Mackey's language) iff $P \leq Q'$. In this case,
we write $P \perp Q$. Mackey at this point imposes his

\vskip 0.406 truecm\par\noindent
{\sl Axiom V$\,$: If $P_{i}$ is any countable family of
pairwise orthogonal elements of $P$, then there exists an element
$P \in L$ with $P_{1} + P_{2} + \cdots  = P$.}
\vskip 0.406 truecm\par\noindent
This axiom guarantees that $L$ is a $\sigma$-{\sl orthomodular}
poset --- that is, $L$ satisfies the two conditions
\begin{itemize}
\item[(a)] Any countable family of pairwise orthogonal elements $P_{i} \in L$ have a join
(least upper bound) $\bigvee_{i} P_{i}$ in $L$, and
\item[(b)] If $P \leq Q$, then $(Q \wedge P') \vee P  = Q$.
\end{itemize}

\minipar

On any such poset $L$, one can define {\sl probability measures}
on $L$ to be mappings $\mu : L \rightarrow [0,1]$ such that
$\mu(1) = 1$ and, for any countable pairwise orthogonal family of
elements $P_{i} \in L$ we have
$\mu(\,{\textstyle\bigvee_{i}} P_{i}) = {\textstyle\sum_{i}}\, \mu(P_{i})$.
We can also define, given any two $\sigma$-OMPs $L$ and $M$, an
$M$-valued measure on $L$ to be a mapping $\alpha : L \rightarrow
M$ such that $\alpha(1_{L}) = 1_{M}$ and, for any countable
pairwise orthogonal family $P_{i} \in L$ we have
$\alpha(\,{\textstyle\bigvee_{i}} P_{i}) = {\textstyle\bigvee_{i}}\alpha(P_{i})$.
For a general discussion of such maps in terms of observables see
[Pt\'ak 2000]. Returning now to the OMP $L$ of questions,
Mackey observes that
\begin{itemize}
\par
\item[(a)] Each state $s \in {\cal S}$ defines a probability measure
$\widehat{s} : L \rightarrow [0,1]$ by evaluation: $\widehat{s}(P_{A,B}) =
P_{A,B}(s) = p_{A}(B|s)$.
\par
\item[(b)] Each observable $A \in {\cal O}$ defines an $L$-valued measure
$P_{A} : {\cal B}({\Bbb R}) \rightarrow L$ via $P_{A}(B) = P_{A,B}$ on the
real Borel sets (which, constituting a $\sigma$-Boolean algebra, certainly
constitute a $\sigma$-OMP).
\end{itemize}
Conversely, suppose $L$ is {\sl any} $\sigma$-OMP $L$, and that
${\cal S}$ is any {\sl order-determining} set of probability
measures on $L$ --- i.e.,
$\mu(p) \leq \mu(q)$ for all $\mu \in {\cal S}$ implies that $p \leq q$.  Let
${\cal O}$ be the set of all $L$-valued Borel measures on the line, and
define $p : {\cal O} \times {\cal S} \rightarrow \Delta({\Bbb R})$ by
$p_{\alpha}(B|\mu) = \mu(\alpha(B))$. Then the structure $({\cal O}, {\cal
S}, p)$ satisfies Mackey's axioms, and, furthermore, the OMP of
questions constructed from it is canonically isomorphic to $L$.

\minipar

As remarked by [Foulis 1962; Gudder 1965], Mackey's axioms then
define the theory of the following class of structures: pairs
$(L,\Delta)$ where $L$ is a $\sigma$-OMP and $\Delta$ is an
order-determining family of probability measures on $L$. Such
pairs (routinely referred to as {\sl quantum logics} in the
mathematical literature in the 1960s and 1970s) have been studied
intensively by many authors. For detailed discussions of orthomodular
posets in the quantum logical context see [Beltrametti and Cassinelli
1981; Gudder 1985; Pt\'ak 2000; Pt\'ak and Pulmannov\'a 1991].
Of course such
quantum logics are still a far cry from the {\sl standard} quantum
logic ${\Bbb P}({\bf H})$. Among other things, the orthomodular poset
${\Bbb P}({\bf H})$ is a {\sl complete lattice}: arbitrary joins exist,
not just countable orthogonal joins. Still, one might hope that a deeper
analysis --- perhaps involving additional axioms --- might lead to
a meaningful characterization, and, ideally, a motivation for the
standard quantum logics.
This was Mackey's expressed goal:
\begin{quote}
``Ideally, one would like to have a list of
physically plausible assumptions from which one could deduce [the
Hilbert space model]. Short of this, one would like a list from
which one could deduce a set of possibilities ..., all but one of
which could be shown to be inconsistent with suitably
planned experiments. At the moment, such lists are not
available...'' [Mackey 1963, p.72]
\end{quote}
This topic lies at the heart of Piron's original
axiomatization, to be discussed in the next section. On the other
hand, as we shall discuss in sections 5 and 6, the autonomous study
of such structures leads naturally to further generalizations,
notably, to orthoalgebras and effect algebras.

\minipar

Before turning to a rapid survey of some of the major developments
which have occurred since Mackey's foundational work, let us make
a few comments. First, and foremost, the major feature which
separates Mackey's formalism from current tendencies in
operational quantum logic is the former's reliance on probability
as a {\sl primitive} concept. While important advances in this
context have been made, for example in [Pulmannov\'a 1986$\,$a,b;
Gudder and Pulmannov\'a 1987; Pulmannov\'a and Gudder 1987], most
contemporary work relegates probability to a derived
notion.\footnote{For example, while in [Piron 1964 \S7]
generalized probability is discussed as a useful physical
heuristic, in [Jauch and Piron 1969
\S5] states are defined as maximal sets of actual properties of
the system. Similarly, while the formalism introduced in [Randall
and Foulis 1970; Foulis and Randall 1972; Randall and Foulis 1973]
is explicitly concerned with operational statistics, in [Foulis,
Piron and Randall 1983; Randall and Foulis 1983; Foulis, Greechie
and R\"uttimann 1992, 1993] emphasis is placed on the conception
of states in terms of supports in the outcome space associated to
the  system.} This is not to say that statistical states are
unimportant in operational quantum logic. However, they have
passed from the status of a rather vaguely construed primitive
concept to that of a well defined structural tool. Here mention
may be made of the characterization of the state spaces of
standard quantum logics, culminating in Navara's proof of the
independence of the automorphism group, center, and state space of
a quantum logic [Navara 1992].

\minipar

A notable exception to this trend is the theory of decision
effects introduced by G\"unther Ludwig during the revision of his
classic text [1954, 1955]. This work is based on the
classification of macroscopic notions into preparative and
effective parts which both participate in measurement interactions
mediated by action carriers. We shall not enter into the details
of Ludwig's axiomatic scheme, successively refined in [Ludwig
1964, 1967, 1968; D\"ahn 1968; Mielnik 1968, 1969; Stolz 1969,
1971; D\"ahn 1972; Ludwig 1972] and codified in the monumental
treatise [Ludwig 1985, 1987], but will content ourselves with some
general remarks. The primitive notion in this theory is that of a
probability relation defined on the Cartesian product of the set
of ensembles and the set of effects, these two sets being taken as
embedded in a suitable pair of Banach spaces. In some sense, then,
the work of Ludwig and his collaborators runs parallel to the
field of operational quantum logic as we have introduced it,
focussing more on the functional analytic structure of the problem
than its ordered algebraic aspects. As such it perhaps bears more
{\sl formal} relation to the algebraic quantum theory of Segal
[1947] and Haag and Kastler [1964] than the operational theories
of Piron and Foulis--Randall to be discussed next. Nevertheless, an
important {\sl physical} feature of Ludwig's work is that it attempts
to deal with the notion of non-ideal measurements by exploiting
subprojective operators. Note that such operators appear naturally
in discussions of generalized localisability [Jauch and Piron
1967; Amrein 1969]. For general surveys of the different
approaches to operational quantum mechanics see [Gudder 1977,
1979, 1981; Ludwig and Neumann 1981], for a detailed analysis
of the model relationship between the approaches of Piron and
Ludwig see [Cattaneo and Laudisa 1994; Cattaneo and Nistic\`o
1993],
and for an overview of the application of POV-measures
to questions in the foundations of quantum mechanics see [Busch,
Lahti and Mittelstaedt 1991; Schroeck 1996].
\par\vskip 0.406 truecm\par
\par\vskip 0.406 truecm\par
\noindent
{\bf 4. The work of Piron}
\par\vskip 0.406 truecm\par
\par\noindent
Significant progress at both ends of the problem of completing and
extending Mackey's programme was made by Constantin Piron [1964]
and further developed in what has become known as the Geneva
School approach to quantum physics. Piron characterized abstractly
those complete orthomodular lattices representable as the lattices
of closed subspaces of {\sl generalized Hilbert spaces}. He also
supplied a deep analysis of the basic physical ideas of quantum
mechanics that helped to motivate the assumptions needed in his
representation theorem as reasonable, general axioms. In this
section we describe a formalized version of these axioms in the
spirit of [Piron 1976], before making some remarks on more recent
developments.
\vfill\eject
\noindent
{4.1. THE REPRESENTATION THEOREM}
\par\vskip 0.406 truecm\par\noindent
The projection lattice ${\Bbb P}({\bf H})$ has a much more regular
structure than the general OMP provided by Mackey's axioms. In
particular, ${\Bbb P}({\bf H})$
\begin{itemize}
\item[(a)] is a {\sl complete} lattice --- that is, the meet and join of
{\sl any} subset of $L$ exist,
\item[(b)] is {\sl atomistic} --- that is, every element of ${\Bbb P}({\bf H})$
is the join of the atoms (here, the one-dimensional projections)
beneath it,
\item[(c)] satisfies the {\sl atomic covering law}: if
$P \in {\Bbb P}({\bf H})$ is an atom and $Q \in {\Bbb P}({\bf H})$
is arbitrary, then $P \vee Q$ covers $Q$, i.e., is an atom in the
lattice $\{M \in L | Q \leq M \}$,
\item[(d)] is {\sl irreducible} --- that is, it cannot be factored as a
non-trivial direct product. Equivalently, no element of
${\Bbb P}({\bf H})$, other than $0$ and $1$, commutes with all
other elements.\footnote{Of course, not every quantum mechanical
system is irreducible, but in general decomposes into a family of
purely quantum systems indexed by {\sl superselection rules}. For
example, [Piron 1964] shows that each orthomodular lattice
satisfying axioms (a)--(c) is the direct union of a family of
irreducible lattices, its corresponding projective geometry being
the direct union of the corresponding geometries. Abstractly,
systems with discrete superselection rules may be treated by
taking projection-valued measures with values in an appropriate
von Neumann algebra $\frak A$. If the induced orthomodular lattice
$L(\frak A)$ does not contain a summand of type $I_2$ then Gleason's
theorem continues to apply: every $\sigma$-additive probability
measure on $L({\frak A})$ extends uniquely to a normal state on
${\frak A}$ [Christensen 1982; Yeadon 1983]. For further discussion see,
for example, [Bunce and Hamhalter 1994; Bunce and Wright 1994;
Hamhalter 1993, 1995].}
\end{itemize}
In his thesis [Piron 1964], Piron proved a partial converse,
namely that all such lattices (of sufficient length) $L$ may be
realised as the set of biorthogonal subspaces of a generalized Hilbert
space. Explicitly, by considering the (essentially) unique meet and atom
preserving embedding of $L$ in a projective geometry, and exploiting the
standard vector space realization of projective geometries of
dimension at least three, he showed that the image of the original
lattice could be characterised by a definite hermitian
form.\footnote{This development has become much more physically
transparent and mathematically elegant since the seminal work of
Faure and Fr\"olicher [1993, 1994, 1995], where the construction
of linear representations for projective geometries and their
morphisms is carried through in a categorically natural manner.
For example, an orthogonality relation determines a morphism from
the projective geometry to its dual and so a quasilinear map from
the underlying vector space to its dual. In this way the inner
product of quantum mechanics finds a rigorous and neat
foundation.}

\minipar

Now, for an arbitrary inner product space $V$, the complete
atomistic ortholattice $L(V)$ of biorthogonal subspaces need not
be orthomodular. When it is, $V$ is termed a {\sl generalized
Hilbert space}. This terminology is motivated by another striking
result, namely that if $V$ is an inner product space over one of
the standard division rings (i.e., ${\Bbb R}, {\Bbb C}$ or ${\Bbb
H}$), then $L(V)$ is orthomodular iff $V$ is complete. This was
first proved by Piron, using a hypothesis on measure extensions
which turned out to be independent of ZF set theory; under
prompting by Stone, a geometric proof was later obtained by
Amemiya and Araki [1965].\footnote{Note that necessary and sufficient
conditions for the underlying division ring to be standard have
recently been found --- one of the simplest statements in the
infinite dimensional case being that the vector space admit an
infinite orthonormal sequence [Sol\`er 1995; Holland 1995; Prestel
1995]; for an example of a nonstandard generalized Hilbert space
see [Keller 1980], for a detailed discussion of the geometry of
generalized Hilbert spaces see [Gross 1979, 1990], and for a
survey of other completeness results see [Dvure\v{c}enskij 1992].}
Finally, let us remark that
the Geneva School formalism that was inspired by this theorem has been
extensively applied to several problems of a more or less concrete
nature, for example, symmetries [Emch and Piron 1962, 1963],
superselection rules [Piron 1965, 1969], observables [Piron 1971;
Giovannini and Piron 1979; Giovannini 1981a,b,c], the {\sl a priori}
probability [Piron 1972], and irreversible processes [Gisin and Piron
1981; Gisin 1981, 1982a,b, 1983a,b].
\par\vskip 0.406 truecm\par
\par\vskip 0.406 truecm\par
\noindent
{4.2. PIRON'S AXIOMS}
\par\vskip 0.406 truecm\par\noindent
Mackey's axioms produce only a $\sigma$-complete orthomodular
poset $L$ --- a far cry from the complete, atomistic OML figuring
in Piron's Theorem. Piron was able to motivate the necessary extra
structure in the context of an axiomatic framework similar to
Mackey's, but differing from it in taking as basic not the concept
of probability, but a concept of {\sl physical property} based on
the {\sl certainty} of obtaining an experimental outcome. Here
Piron consciously exploits the work of Dirac [1930 \S1.2], who
gives an operational discussion of light polarisation in terms of
the certainty or otherwise of passage through an appropriate
crystal, and the conception of Einstein, Podolsky and Rosen [1935]
that elements of reality are sufficient conditions that one be
able to predict a physical quantity with certainty and without
disturbing the system.

\minipar

Piron begins with a primitive set ${\cal Q}$ of {\sl questions}
--- understood to represent definite experimental projects having
just two possible outcomes, which we designate as {\sl yes} and
{\sl no}. For ease of presentation let us consider given a set
${\cal P}$ of {\sl preparation procedures}.\footnote{Note that
this is not strictly necessary, but is just an expedient to avoid
locutions such as `if the system is, or has been prepared, in such
a way that $\ldots$'. Similarly, the usual identification of
propositions with equivalence classes of questions is made for ease of
exposition and should not be taken too seriously as a definition.}
For $P \in {\cal P}$ and $\alpha \in {\cal Q}$ we write $P \vDash
\alpha$ to indicate that the preparation $P$ is such that the
answer to the question $\alpha$ can be predicted with certainty to
be {\sl yes}. We can then associate, to every question $\alpha$,
the {\sl proposition}
$$[\alpha] = \{~ P \in {\cal P} ~|~ P \vDash \alpha~\}\,.$$ Let
${\cal L} := \{~ [\alpha] ~|~ \alpha \in {\cal Q}~\}$
be the set of all such propositions, considered as a poset under
set inclusion. Note that $[\alpha] \subseteq [\beta]$ iff every
preparation making $\alpha$ certain also makes $\beta$ certain.
Piron proceeds to adduce several axioms the force of which is to
make ${\cal L}$ a complete, atomistic OML satisfying the covering
law.

\vskip 0.406 truecm\par
\noindent{\bf ${\cal L}$ is a complete lattice}  The first, and
probably the most novel, of these axioms involves the notion of a
{\sl product question}. Given a non-empty set $A$ of questions,
their product is the question $\alpha = \Pi A$ defined as follows:
to pose $\alpha$, one selects, in any way one will, a question
$\beta \in A$ and, posing this question, attributes to $\alpha$ the
answer obtained. Piron's first axiom requires that ${\cal Q}$ be
closed under the formation of arbitrary product questions. A
moment's reflection reveals that $[\Pi A] = \bigcap_{\beta \in A}
[\beta]$. Hence, ${\cal L}$ is closed under arbitrary
intersections and thus a complete lattice.\footnote{The product
operation was first introduced in [Jauch and Piron 1969]. Earlier,
the meet had been introduced via either semantic conjunction
[Piron 1964] or limit filters [Jauch 1968].}

\vskip 0.406 truecm\par
\noindent{\bf Orthocomplementation}
If $\alpha$ is any question, we may define an {\sl inverse}
question $\alpha^{\sim}$ by interchanging the roles of {\sl yes}
and {\sl no}. Piron requires that ${\cal Q}$ be closed under the
formation of inverses. The intended interpretation requires us to
suppose that $[\alpha] \cap [\alpha^{\sim}] = \emptyset$. In order
to secure an orthocomplementation on ${\cal L}$, Piron introduces
another axiom, namely, that for any question $\alpha$, there
exists some {\sl compatible complement} $\beta \in [\alpha]$
satisfying $[\beta^{\sim}] \vee [\alpha] = 1$.\footnote{The fact
that an axiom must be postulated guaranteeing the existence of an
orthocomplementation is due to the fact that the inverses of
equivalent questions need not themselves be equivalent. For
example, $0\cdot I = 0$ however
$(0 \cdot I)^{\sim} = 0^{\sim} \cdot I^{\sim} = I \cdot 0 = 0$
and $0^{\sim} = I$. For a discussion of some confusions on this
point see [Foulis and Randall 1984].}

\vskip 0.406 truecm\par
\noindent{\bf Orthomodularity}
On the face of it, this does not rule out the possibility that
there may exist several inequivalent compatible complements for a
given $\alpha$. However, this is remedied by a third axiom,
Piron's

\vskip 0.406 truecm\par\noindent
{\sl Axiom P$\,$: If $b < c$ and $b'$ and $c'$ are compatible
complements for $b$ and $c$, respectively, then the sub-lattice of
${\cal L}$ generated by ${b,b',c,c'}$ is distributive.}
\vskip 0.406 truecm\par\noindent
It follows that compatible complements are unique, thereby
defining an orthocomplementation. Moreover, Axiom P dictates that
${\cal L}$ be orthomodular: if $b < c$, then
$(c \wedge b') \vee b = c$ by the distributivity of $\{b,c,b',c'\}$.

\vskip 0.406 truecm\par
\noindent {\bf Atomicity and the Covering Law} Piron enforces
the atomicity of the lattice with an ad-hoc axiom (A1) requiring
that $L$ be atomic --- i.e., every element dominates at least one
atom. The covering law is also imposed directly (as axiom A2), but
with some substantial motivation, as follows. Let us denote by
$\Sigma_{\scriptscriptstyle L}$ the set of atoms of $L$. Now, in
any orthocomplemented poset $L$, the {\sl Sasaki mapping}
$\phi : L \times L \rightarrow L$ is given by $\phi(a,b) = b
\wedge (b' \vee a)$. If $b$ is fixed, we write $\phi_{b} : L
\rightarrow L$ for the mapping $\phi_{b}(a) = \phi(a,b) = b \wedge
(b' \vee a)$. Note that $L$ is orthomodular iff $\phi_{b}(a) = a$
for all $a \leq b$, in which case $\phi_{b}(a) \vee b' = a
\vee b'$. Using these remarks, it is not hard to prove that an OML $L$
satisfies the atomic covering law iff, for all $a \in L$, we have
$a \in \Sigma_{\scriptscriptstyle L} \& b \not\perp a
\Rightarrow  \phi_{b}(a) \in \Sigma_{\scriptscriptstyle L}$.
We then have an alternative formulation of the covering law,
namely, that Sasaki projections map atoms either to atoms, or to
$0$.

\minipar

Piron defines the {\sl state} of the system to be the set of all
propositions $p = [\alpha]$ that are certain (at a given time, in
a given situation).  We naturally require that the state be closed
under intersection and enlargement, i.e., that it be a complete
filter in the lattice ${\cal L}$. Such a filter is principal, and
is generated by an atom. Hence, states may be represented by
atoms.\footnote{Conversely, if $p \not = 0$, then $p = [\alpha]$
where $\alpha$ is a question that is certain for at least one
preparation. Hence, there exists at least one state (i.e., any
state compatible with that preparation) that contains $p$. Hence,
there must be sufficiently many states so that for every $p \in
{\cal L}$, there is a state/atom $a \leq p$. But this easily
implies that every atom is a state. Thus, states correspond {\sl
exactly} to atoms of ${\cal L}$.} Finally, call $a, b$ in ${\cal
L}$ {\sl compatible} iff
$\{a,b,a',b'\}$ is distributive, iff $\phi_{b}(a) = a \wedge b$.
Piron calls the question
\begin{itemize}
\item[(a)]  {\sl ideal} iff every proposition compatible with
$[\alpha]$ that is certain before a measurement of $\alpha$ is
also certain again afterwards when the result of that measurement
is {\sl yes}.
\item[(b)] {\sl first-kind} iff the answer to $\alpha$ immediately after
securing the answer {\sl yes} is {\sl certain} to be again {\sl yes}.
\end{itemize}
In the presence of axiom A2 (i.e., the covering law), one can then
prove that, for $\beta$ an ideal first-kind measurement of $b
\in {\cal L}$, if $a$ is the state before measurement, then the
state after securing {\sl yes} upon measurement of $\alpha$ is
$\phi_{b}(a)$.

\minipar

Many people have found the physical reasoning that motivates
Piron's axioms compelling. However, this framework turns out to
have some sharp limitations. In particular, a system consisting of
two ``separated'' systems, in the sense of Aerts [1981, 1982],
each of which individually obeys Piron's axioms, will as a whole
conform to these axioms if and only if one of the systems is
classical. To prove this key result, Aerts exploited the notion of
an {\sl orthogonality relation}, where two states are orthogonal
if there exists a question which is certain for the first and
impossible for the second. The use of this relation has become
central in more recent axiomatizations of the Geneva School
approach, such as [Piron 1990; Moore 1999]. For a
detailed analysis see [Valckenborgh 2000].
Note that in these
works attention is focussed on complete atomistic ortholattices as
models for the most direct axiomatizations based on the physical
duality between the state and property descriptions of a physical
system. Somewhat paradoxically, then, Piron's approach keeps one
axiom rejected in the OMP approach
--- namely completeness --- and rejects another which the latter keeps
--- namely (some form of) orthomodularity.
As we shall see in the next section, this cleavage is symptomatic of
the fact that one should distinguish conceptually the property lattice
of a system from its logic, even when they turn out to be isomorphic.

\vskip 0.406 truecm\par
\par\vskip 0.406 truecm\par
\noindent
{\bf 5. The work of Foulis and Randall}
\par\vskip 0.406 truecm\par
\par\noindent
Contemporary with these developments was the work of Dave Foulis
and the late Charlie Randall on {\sl empirical logic}, a happy
synthesis of ideas coming from their respective doctoral
dissertations, on abstract lattice theory [Foulis 1958] and
concrete operational statistics [Randall 1966]. Not only does this
formalism provide a powerful general heuristic, but as we shall
see it has also laid the ground for several of the purely
mathematical developments to be discussed in the following.
\par\vskip 0.406 truecm\par
\par\vskip 0.406 truecm\par
\noindent
{5.1. TEST SPACES}
\par\vskip 0.406 truecm\par\noindent
Both Mackey and Piron begin with a primitive structure in which
experimental propositions of the form ``observable $A$ takes value
in set $B$'' are unrelated for distinct observables. In effect,
each observable $A$ is associated with a Boolean algebra ${\cal
B}_{A}$ of possible events (isomorphic to the Borel field in
Mackey's scheme, and to $\{0,1\}$ in Piron's), these Boolean algebras
being initially disjoint from one another. Identifications are then made
between the Boolean algebras corresponding to different
observables.  In Mackey's scheme, the primitive propositions
$(A_{1},B_{1})$ and $(A_{2},B_{2})$ are identified iff they are {\sl
equiprobable} in every state; in Piron's, iff they are {\sl
certain} in exactly the same situations. Both approaches to the
construction of quantum logics have been the object of some
criticism. In particular, as a number of authors point out, both
become problematic when one considers compound or iterated
measurements.\footnote{A simple example is offered in [Cooke and
Hilgevoord 1981]. The point is a familiar one --- even in orthodox
Hilbert space quantum mechanics, one must keep track of phase
relations in discussing iterated experiments, and these are lost
when one identifies experimental propositions according to either
the Mackey or the Piron scheme.}

\minipar

In a long series of papers (e.g., [Foulis and Randall 1972, 1974,
1978, 1981a; Randall and Foulis 1970, 1973, 1978, 1983a]), Foulis
and Randall developed an extensive theory -- which they termed
{\sl empirical logic} --- in which such identifications are given
{\sl a priori}, with no prior reference to any concept of state or
property. Their formalism is based on the primitive notion of an
{\sl operation} or {\sl test} --- that is, a definite set of
mutually exclusive alternative possible {\sl outcomes}. The
Foulis-Randall theory focusses on {\sl test spaces}, i.e.,
collections $\frak A$ of overlapping tests. The identification of
outcomes between distinct tests is understood to be given, i.e.,
Foulis and Randall lay down no doctrine as to how such
identifications must be made. Letting
$X=\bigcup\frak A$ stand for the {\sl outcome space} of
$\frak A$, a {\sl statistical state} on $\frak A$ is defined to be
a mapping $\omega : X \rightarrow [0,1]$ such that $\sum_{x \in E}
\omega(x) = 1$ for every test $E \in {\frak A}$, and a {\sl
realistic state} is represented [Foulis, Piron and Randall 1983]
by a certain kind of subset of $X$ called a {\sl support},
representing the totality of outcomes possible in that state. Note
that, quite apart from its own merits, this notion can be used to
give a perspicuous mathematical treatment of Piron's axiomatics;
see [Randall and Foulis 1983b] and [Wilce 1997].

\minipar

A number of algebraic, analytic and order-theoretic objects can be
attached to a test space $\frak A$, each serving in a slightly
different way as a sort of ``logic''. Under simple normative
conditions on the combinatorial  structure of $\frak A$, these
turn out to coincide with more familiar structures. In particular,
if $\frak A$ is ``algebraic'',\footnote{A test space is algebraic if any two elements sharing a common
complement share exactly the same complements, where $A$ and $B$
are complements if they are disjoint and $A\cup B$ is a test.}
one can construct from the events
of $\frak A$ a rather well-behaved ordered partial algebraic
structure $\Pi({\frak A})$, called an {\sl orthoalgebra}. These
can be defined abstractly: an orthoalgebra is a pair $(L,\oplus)$
where $L$ is a set and $\oplus$ is a commutative, associative {\sl
partial} binary operation on $L$ satisfying the three additional
conditions:
\begin{itemize} \item[(a)] There exists a neutral element $0 \in L$ such that,
for every $p \in L$, $p \oplus 0  = p$,
\item[(b)] There exists a unit element $1 \in L$ such that, for every $p \in
L$, there is a unique $q \in L$ with $p \oplus q = 1$,
\item[(c)] If $p \oplus p$ exists, then $p = 0$.
\end{itemize}
Orthoalgebras then generalize orthomodular posets, which can be
defined as orthoalgebras in which, given that $p \oplus q$,
$q \oplus r$ and $r \oplus p$ all exist, the element
$p \oplus q \oplus r$ also exists. This axiom, called
{\sl orthocoherence}, is in fact a finitistic version of Mackey's
axiom V. Conversely, dropping condition (c) we obtain what is
called an {\sl effect algebra}, called generalized orthoalgebras
by  [Giuntini and Greuling 1989] and D-posets by [K\^opka 1992].
\par\vskip 0.406 truecm\par
\par\vskip 0.406 truecm\par
\noindent
{5.2. ORTHOALGEBRAS}
\par\vskip 0.406 truecm\par\noindent
Orthoalgebras and effect algebras are sufficiently regular objects
to have an interesting mathematical theory (one that is only
beginning to be explored). In particular, nearly all of the
conceptual apparatus of OMP-based quantum logic, such as centers
[Greechie, Foulis and Pulmannov\'a 1995] and Sasaki projections
[Bennett and Foulis 1998; Wilce 2000], can be rather easily
extended to this more general context. On the other hand, because
of their simplicity, test spaces are often much easier to
manipulate than their associated ``logics''. They also have the
heuristic advantage that the operational interpretation is, so to
say, right on the surface, with the logics serving only as useful
invariants. In particular, while it is completely straightforward
to combine test spaces sequentially, the various ``logics'' rarely
respect such combinations. Finally, if $\frak A$ is algebraic,
there exists a canonical order-preserving mapping
$L \rightarrow {\cal L}$ from the logic of $\frak A$ into the
property lattice associated with any entity
$({\frak A},\Sigma)$ over ${\frak A}$.  In both classical and
quantum mechanical examples, this mapping is in fact an
isomorphism, so that $L$ inherits from ${\cal L}$ the structure of
a complete lattice, and
${\cal L}$ inherits from $L$ an orthocomplementation and
orthomodularity. This isomorphism is, however, the exception
rather than the rule. As stressed by [Foulis, Piron and Randall
1983], the tendency to identify ${\cal L}$ and $L$ -- even when
they are isomorphic --- has caused a great deal of unnecessary
confusion in discussions of the foundations and interpretation of
quantum mechanics.

\minipar

Besides its awkwardness in dealing with sequential measurements,
another difficulty that arises with the Mackey scheme of quantum
logic, again recognized first by Foulis and Randall [1979], is
that it is not stable under the formation of any reasonable sort
of {\sl tensor product}. Given quantum logics $(L,\Delta)$ and
$(L',\Delta')$, each understood to represent some ``physical'' system,
one wants to construct a model $(M, \Gamma)$ of the coupled system
in which $L$ and $L'$ may display correlations, but do not
directly interact. Minimal requirements would be that
\begin{itemize}
\item[(a)] there exists a map $L \times L' \rightarrow M$ carrying $p, q$ to
some representative proposition $p \otimes q \in M$, and
\item[(b)] for every pair of states $\mu \in \Delta, \nu \in \Delta'$, we be
able to form a state $\mu \otimes \nu \in \Gamma$ such that $(\mu
\otimes
\nu)(p \otimes q)  = \mu(p) \nu(q)$.
\end{itemize}
However, Foulis and Randall produce a simple example showing that
this is in general impossible: a small, finite OML $L$ with a full
set of states such that no such ``tensor product'' exists for two
copies of $L$.

\minipar

The culprit turns out to be Mackey's Axiom V --- or, more
precisely, orthocoherence. Indeed, one can show, under the very
mild assumption that the orthoalgebras involved each carry a
unital family of states, that tensor products of orthoalgebras
can be formed in such a way that desiderata (a) and (b) are
satisfied [Foulis and Randall 1981b; Randall and Foulis 1981].
However, as the example just discussed illustrates, orthocoherence
is not stable under this tensor product. Combining these results
with the above mentioned negative results of Aerts for property
lattices, what emerges is that the isomorphism between logic and
property lattice, characteristic of both quantum and classical
systems, breaks down when one forms tensor products unless the
systems in question are classical. This is not to say that the
results were entirely negative. Later research into the structure
of tensor products [Kl\"ay, Randall and Foulis 1987; Golfin 1987;
Wilce 1990, 1992; Bennett and Foulis 1993; Dvure\v{c}enskij and
Pulmannov\'a 1994; Dvure\v{c}enskij 1995] revealed that
Foulis--Randall tensor products of quantum-mechanical entities,
while no longer strictly quantum, still retain a rich geometric
structure.

\minipar

These results gave substantial impetus to the study of
orthoalgebras, test spaces and other structures more general than
those considered by Mackey and Piron (some of which will be
discussed below). The theory of test spaces, in particular, has
developed in several directions in the past decade.  A number of
authors (e.g., [Dvure\v{c}enskij and Pulmannov\'a 1994b;
Pulmannov\'a and Wilce 1995; Gudder 1997]) have discussed
generalized test spaces in which outcomes are permitted to occur
with some multiplicity or intensity, and have used these to
provide an operational semantics for effect algebras that
parallels the test-space semantics for orthoalgebras. Measure
theory on orthoalgebras has been discussed by [Habil 1993]. Nishimura
[1993, 1995] has generalized the idea of a test space by replacing
discrete outcome sets by complete Boolean algebras and locales.  [Wilce
2000] gives an up-to-date survey of the Foulis-Randall
theory; for a personal view of the historical development of this
strand of operational quantum logic see [Foulis 1998, 1999].
\par\vskip 0.366 truecm\par
\par\vskip 0.386 truecm\par
\noindent
{\bf 6. Orthomodular structures}
\par\vskip 0.486 truecm\par
\noindent
Thus far, we have focussed on quantum logic as a foundational or
interpretive programme in physics. But the subject has other,
quite independent roots in pure mathematics. Von Neumann himself
had stressed the importance of order theoretic methods in studying
infinite-dimensional analogues of projective geometry.  Loomis
[1955] and Maeda [1955] independently recognized that a fair bit
of the dimension theory of von Neumann algebras could be carried
over into a purely lattice-theoretic setting, namely, that of an
orthomodular lattice equipped with a suitable equivalence
relation. This stimulated a number of mathematicians to begin
investigating orthomodular lattices {\sl in abstracto}. It soon
became apparent that such lattices occur with some naturality in a
wide range of mathematical contexts.
If $(S,*)$ is any involutive semigroup, call an element $e
\in S$ satisfying $p = p^{2} = p^{*}$ a {\sl projection}. If
$S$ contains a (two-sided) zero element, the {\sl right
annihilator} of $x \in S$ is the right ideal
$\{\,a\in S\,|\,ax=0\,\}$. Foulis [1958, 1960, 1962] defined
a {\sl Baer *-semigroup} to be an involutive semigroup $S$ with
zero having the property that the right-annihilator of any element
$x \in S$ is the right ideal generated by a (necessarily, unique)
projection $x'$. He showed that the set $L(S)$ of {\sl closed}
projections $p = p''$ in $S$ always forms an orthomodular lattice.
Conversely, every orthomodular lattice can be represented as
$L(S)$ for some Baer *-semigroup. Indeed, while this
representation is not unique, there is a canonical choice for $S$,
namely the semigroup $S(L)$ of residuated self-mappings of $L$,
i.e., mappings $\phi : L
\rightarrow L$ for which there exists a mapping
$\psi : L \rightarrow L$ satisfying
$\psi(x) \leq y' \ \Leftrightarrow \ x \leq \phi(y)'\,$.
Among these mappings are the Sasaki projections $\phi_{b}$,
discussed in section 4, which turn out to be exactly
the closed projections in $S(L)$.

\minipar

Over the following decades, a substantial pure theory of
orthomodular lattices was developed by Foulis and others. The
state of this theory as of the early 1980s is represented by the
book of Kalmbach [1983]. [Bruns and Harding 2000] discuss more
recent developments, of which there have been many. Particularly
striking is the recent discovery of Harding [1996, 1998] that one
can organize the set of direct-product decompositions of
essentially any algebraic object into an orthomodular poset. On
the other hand, the on-going work on Mackey's programme also
produced a variety of structures more general than orthomodular
lattices and posets --- orthoalgebras, the still more general
effect algebras, and, in a different direction, the partial
Boolean algebras of Kochen and Specker
[1967]. All of these are primarily partial algebraic, and only
secondarily order theoretic, objects. All have attracted,
especially during the past few years, significant mathematical
interest.\footnote{One notable result is that of Kochen and Conway
[Kochen 1996], that very small sets of projections in ${\Bbb
P}({\bf H})$ generate a partial Boolean algebra that is dense in
the full projection lattice.} The theory of effect algebras, much
of which is due to the pioneering work of Foulis and the late
M.~K.~Bennett [Bennett and Foulis, 1995, 1997; Foulis and Bennett
1994], continues to develop rapidly. Of particular interest here
is their recent reformulation of a large part of the theory of
effect algebras (and thus, of quantum logic) as a branch of the
theory of ordered abelian groups, also discussed in [Foulis,
Bennett and Greechie 1996; Foulis, Greechie and Bennett 1998;
Wilce 1995, 1998]. This is the subject of [Foulis 2000].

\minipar

Finally, orthomodular lattices have also been studied in detail in
the purely logical context, and in particular the possibility of
defining reasonable implication connectives. One of the basic
results in this direction is that of Kalmbach [1974] who,
exploiting the characterisation of free orthomodular lattices on
two generators [Bruns and Kalmbach 1973], was able to show that
there are exactly five lattice polynomials $a\rightarrow b$
satisfying the primitive implicative condition
$a\leq b\Leftrightarrow(a\rightarrow b)=1$. Note that here
orthomodularity is essential, a simple consideration of the
non-orthomodular ``benzene ring'' showing that such connectives do
not exist in the non-orthomodular case [Moore 1993]. For an
analysis of the weaker exportation condition $a\leq
b\Rightarrow(a\rightarrow b)=1$ conjoined with {\sl modus ponens}
see [Herman, Marsden and Piziak 1975], and for a detailed
investigation of the deduction theorem see Malinowski [1990,
1992]. On the other hand, defining a Kripkean accessibility
relation induced from {\sl non}-orthogonality, an idea having its
origins in Foulis and Randall's work on lexicographic
orthogonality [1971], has allowed the introduction of modal
quantum logic [Dalla Chiara 1977, 1983; Goldblatt 1974, 1975]. Of
course there has been much other work on implications in quantum
logic; for general overviews see, for example, [Dalla Chiara 1986;
van Fraassen 1981; Hardegree and Frazer 1981].

\vskip 0.406 truecm\par
\par\vskip 0.406 truecm\par
\noindent
{\bf 7. Dynamical, categorical and computational aspects}
\par\vskip 0.406 truecm\par
\par\noindent
We close by considering the categorical reformulation of the basic
notions of order structures and its application to operational
quantum theory, a subject with strong links to various recent
developments in enriched category theory and computational
semantics. The basic tool of this theory is that of
pairs $f\dashv g\,$, where $f:L\rightarrow M$ and $g:M\rightarrow L$ are isotone
maps between posets, satisfying the
adjunction condition
$f(a)\leq b\Leftrightarrow a\leq g(b)\,$.
For a pedestrian development of the theory of adjunctions with a
particular focus on its operational applications we refer to [Coecke
and Moore 2000]. It is amusing to note that this notion may be
used to shed some light on Birkhoff and von Neumann's remark cited
above that while philosophers have tended to focus on the nature of
negation in non-classical logics, the study of quantum mechanics
highlights the distributive law as the weak link in  operational
quantum logic. To see this, let us note that Heyting algebras,
considered as models for intuitionistic logic, may be defined as those
lattices admitting an implication connection $\rightarrow$ satisfying
the adjunction condition
$(x\wedge a)\leq b\Leftrightarrow x\leq(a\rightarrow b)$
[Birkhoff 1940
\S161; Birkhoff 1942 \S27]. Since the condition $f\dashv g$ implies that
$f$ preserves existing joins and $g$ preserves existing meets, any
Heyting algebra is distributive. On the other hand, much of the
structure theory of orthomodular lattices rests on the so-called
Sasaki adjunction $\phi_a\dashv\phi^a$, where
$$\phi_a(x)=a\wedge(a'\vee x)\ \ {\rm and}\ \ \phi^a(x)=a'\vee(a\wedge
x)$$ [Nakamura 1957; Sasaki 1955]. In a certain sense, then, we may
consider Heyting algebras as a class of distributive lattices
where those elements possessing a complement may be simply
characterised, and orthomodular lattices as a class of
ortholattices where the set of complements of any given element
may be simply computed. For discussions of Heyting algebras and
the more general semicomplemented lattices, see [Curry 1963; Frink
1962; K\"ohler 1981; Nemitz 1965].

\minipar

One of the earliest, and most important, applications of
residuations in orthomodular lattices was Foulis' pioneering work
on Baer $^\ast$-semigroups, described above. This research has not
only led to a deeper understanding of the notion of residuation
[Blyth and Janowitz 1972; Derd\'erian 1967], but was also crucial
in the development of dynamical aspects of operational quantum
logic. One of the first of these developments was the work of Pool
[1968a,b], who sought a phenomenological interpretation of Baer
$^{*}$-semigroups via the notion of conditional probability
supplied by the conventional quantum theory of measurement. A more
distinctly operational approach to evolutions in general was
provided by Daniel [1982, 1989] and extended by Faure, Moore and
Piron [1995], the latter leading to a general study of the
categories of state spaces and property lattices [Moore 1995,
1997]. Here an externally imposed evolution is modeled by pulling
back definite experimental projects defined at the final time to
their images defined at the initial time. By physical arguments,
this map must preserve the product operation and so the lattice
meet. Hence, under suitable stability conditions, its join
preserving left adjoint then describes the propagation of the
state of the system. These observations have been generalized by
Amira, Coecke and Stubbe [1998], who explicate the structure of
operational tests derived from the notions of free choice and
composition. Note that the latter of these notions in particular
plays a fundamental role in the heuristics of Foulis and Randall
mentioned above. Finally, the abstract structure of operational
resolutions has been analysed by Coecke and Stubbe [1999a,b,
2000], allowing, for example, an analysis of the physical notions
of compoundness [Coecke 2000] and the duality between causality
and propagation [Coecke, Moore and Stubbe 2000].

\minipar

Mathematically, the structure induced from operational resolutions
is that of a {\sl quantaloid}, namely a category whose Hom-sets
are join complete lattices such that composition distributes on
both sides over joins.\footnote{The name quantaloid was introduced
by Rosenthal [1991], although much of the basic conceptual
development had already been made by Joyal and Tierney [1984] and
Pitts [1988] in their studies of Grothendieck topoi.} One thus
obtains a simple example of an {\sl enriched category}, in which
Hom-sets are objects in some base category and composition is
realised by natural transformations satisfying coherence criteria.
This has become a central notion in category theory, and is
treated in the standard text [Borceux 1994]; for a specialised
treatment see [Kelly 1982] and for a pedagogical development see
[Borceux and Stubbe 2000]. Restricting attention to categories
with a single object, we then recover {\sl quantales}, introduced by
Mulvey [1986] as a non-commutative generalization of
locales.\footnote{Note that historical precedents for these notions can
be traced back to the work of Ward and Dilworth [Ward 1937, 1938; Ward
and Dilworth 1939a,b], who used such multiplications to study ideals in
rings, a technique which has recently been applied to non-commutative
C$^\ast$-algebras [Borceux, Rosick\'y and Van Den Bossche 1989;
Rosick\'y 1989].} Here mention may be made of the recent extension of
the localic notions of simplicity and spatiality to the context of
quantales [Kruml 2000; Paseka 1997; Paseka and Kruml 2000; Rosick\'y
1995], a subject treated in some detail in [Paseka and
Rosick\'y 2000].
It is interesting to observe that similar structures have also
been exploited in computer science. An important example is the
so-called {\sl observational logic} of Abramsky and Vickers [Abramsky
1991; Abramsky and Vickers 1993; Vickers 1989]. Here it is
observed that the possibility that observation induces a change of
state formally leads to a passage from frames to quantales. This
line of thought has been extended by Resende [1999, 2000], who
describes general systems on the basis of their observable
behaviour independently of any supposed state space. For an overview see the
[Resende 2000].  It should be
noted, however, that these considerations are rather different
from the contemporary notion of quantum computation (see for
example [Gudder 2000]'s general theory of quantum languages, i.e.,
languages accepted by quantum automata).

\vskip 0.406 truecm\par
\par\vskip 0.406 truecm\par
\noindent
{\Large{\bf References}}
\par\vskip 0.256 truecm\par
\par\noindent

\newcount\numref
\numref=0
\begin{itemize}
\def\refjl#1#2#3#4#5#6{\advance\numref by1\item[{[\the\numref]}]
\par\vskip-4pt\par#1 (#2) #3, {\it #4} {\bf #5}, #6.}
\def\refbk#1#2#3#4{\advance\numref by1\item[{[\the\numref]}]
\par\vskip-4pt\par#1 (#2) {\sl #3}, #4.}
\def\refpr#1#2#3#4#5#6{\advance\numref by1\item[{[\the\numref]}]
\par\vskip-4pt\par#1 (#2) #3, in #4 {\sl #5}, #6.}
\refjl{Abramsky, S.}{1991}{Domain theory in logical form}{Annals of Pure and
Applied Logic}{51}{1--77}
\par\vspace{-0.08cm}\par\noindent\refjl{Abramsky, S. and Vickers, S.}{1993}{Quantales, observational logic, and
process semantics}{Mathematical Structures in Computer
Science}{3}{161--227}
\par\vspace{-0.08cm}\par\noindent\refbk{Aerts, D.}{1981}{The One and the Many$\,$: Towards a Unification of the
Quantum and the Classical Description of One and many Physical
Entities}{Doctoral Dissertation, Free University of Brussels}
\par\vspace{-0.08cm}\par\noindent\refjl{Aerts, D.}{1982}{Description of many  separated physical entities
without the paradoxes encountered in quantum
mechanics}{Foundations of Physics}{12}{1131--1170}
\par\vspace{-0.08cm}\par\noindent\refjl{Amemiya, I. and Araki, H.}{1967}{A remark on Piron's paper}{Publications
of the Research Institute of Mathematical Sciences Kyoto
University {\rm A}}{2}{423--429}
\par\vspace{-0.08cm}\par\noindent\refjl{Amira, H., Coecke, B., and Stubbe, I.}{1998}{How quantales emerge by
introducing induction within the operational approach}{Helvetica
Physica Acta}{71}{554--572}
\par\vspace{-0.08cm}\par\noindent\refjl{Amrein, W.O.}{1969}{Localizability for particles of mass zero}{Helvetica
Physica Acta}{42}{149--190}
\par\vspace{-0.08cm}\par\noindent\refbk{Bamberg, P., and Sternberg, S.}{1990}{A Course in Mathematics for Students of
Physics}{Harvard University Press, Cambridge}
\par\vspace{-0.08cm}\par\noindent\refbk{Beltrametti, E.G. and Cassinelli, J.}{1981}{The Logic of Quantum
Mechanics}{Addison--Wesley, Reading}
\par\vspace{-0.08cm}\par\noindent\refjl{Bennett, M.K. and Foulis, D.J.}{1995}{Phi--symmetric effect
algebras}{Foundations of Physics}{25}{1699--1722}
\par\vspace{-0.08cm}\par\noindent\refjl{Bennett, M.K. and Foulis, D.J.}{1997}{Interval algebras and unsharp
quantum logics}{Advances in Mathematics}{19}{200--215}
\par\vspace{-0.08cm}\par\noindent\refjl{Bennett, M.K. and Foulis, D.J.}{1998}{A generalized Sasaki projection
for effect algebras}{Tatra Mountains Mathematical
Publications}{15}{55--66}
\par\vspace{-0.08cm}\par\noindent\refbk{Blyth, T.S. and Janowitz, M.F.}{1972}{Residuation Theory}{Pergamon
Press, New York}
\par\vspace{-0.08cm}\par\noindent\refbk{Birkhoff, G.}{1940}{Lattice Theory}{American Mathematical
Society, Providence}
\par\vspace{-0.08cm}\par\noindent\refjl{Birkhoff, G.}{1942}{Lattice--ordered groups}{Annals of
Mathematics}{43}{298--331}
\par\vspace{-0.08cm}\par\noindent\refjl{Birkhoff, G. and von Neumann, J.}{1936}{The logic of quantum
mechanics}{Annals of Mathematics}{37}{823--843}
\par\vspace{-0.08cm}\par\noindent\refbk{Borceux, F.}{1994}{Handbook of Categorical Algebra}{Cambridge University
Press, Cambridge}
\par\vspace{-0.08cm}\par\noindent\refjl{Borceux, F., Rosick\'y, J. and Van Den Bossche, G.}{1989}{Quantales and
C$^\ast$--algebras}{Journal of the London Mathematical
Society}{40}{398--404}
\par\vspace{-0.08cm}\par\noindent\refpr{Borceux, F. and Stubbe,
I.}{2000}{Short introduction to enriched categories}{B. Coecke, D.J.
Moore and A. Wilce (eds.)}{Current Research in Operational
Quantum Logic: Algebras Categories, Languages}{Kluwer Academic
Publishers}
\par\vspace{-0.08cm}\par\noindent\refpr{Bruns, G. and Harding,
J.}{2000}{Algebraic aspects of orthomodular lattices}{B. Coecke, D.J.
Moore and A. Wilce (eds.)}{Current Research in Operational
Quantum Logic: Algebras Categories, Languages}{Kluwer Academic
Publishers}
\par\vspace{-0.08cm}\par\noindent\refpr{Bruns, G. and Kalmbach, G.}{1973}{Some remarks on free
orthomodular lattices}{J. Schmidt (ed.)}{Proceedings of the
Lattice Theory Conference}{Houston}
\par\vspace{-0.08cm}\par\noindent\refjl{Bunce, L.J. and Hamhalter, J.}{1994}
{Jauch--Piron states on von
Neumann algebras}{Mathematische Zeitschrift}{215}{491--502}
\par\vspace{-0.08cm}\par\noindent\refjl{Bunce, L.J. and Wright, J.D.M.}
{1994}{The Mackey--Gleason
problem for vector measures on projections in von Neumann
algebras}{Journal of the London Mathematical
Society}{49}{133--149}
\par\vspace{-0.08cm}\par\noindent\refbk{Busch, P., Lahti, P.J. and
Mittelstaedt, P.}{1991}{The Quantum Theory of
Measurement}{Springer-Verlag (Berlin)}
\par\vspace{-0.08cm}\par\noindent\refjl{Cattaneo, G. and Laudisa,
F.}{1994}{Axiomatic unsharp quantum theory (from Mackey to Ludwig and
Piron)}{Foundations of Physics}{24}{631--683}
\par\vspace{-0.08cm}\par\noindent\refjl{Cattaneo, G. and Nistic\`o, G.}{1993}{A model of Piron's
preparation--question structures in Ludwig's selection
structures}{International Journal of Theoretical
Physics}{32}{407--431}
\par\vspace{-0.08cm}\par\noindent\refjl{Christensen, E.}{1982}{Measures on projections and physical
states}{Communications in Mathematical Physics}{86}{529--538}
\par\vspace{-0.08cm}\par\noindent\refjl{Coecke, B.}{2000}{Structural characterization of compoundness}{International
Journal of Theoretical Physics}{39}{581--590}
\par\vspace{-0.08cm}\par\noindent\refpr{Coecke, B. and Moore,
D.J.}{2000}{Operational Galois adjunctions}{B. Coecke, D.J.
Moore and A. Wilce (eds.)}{Current Research in Operational
Quantum Logic: Algebras Categories, Languages}{Kluwer Academic
Publishers}
\par\vspace{-0.08cm}\par\noindent\refjl{Coecke, B., Moore, D.J. and Stubbe, I.}{2001}{Quantaloids
describing causation and propagation for physical properties}{Foundations of Physics
Letters}{14}{133--145}
\par\vspace{-0.08cm}\par\noindent\refbk{Coecke, B., Moore, D.J. and
Wilce, A. [Editors]}{2000}{Current Research in Operational Quantum
Logic: Algebras, Categories, Languages}{Kluwer Academic Publishers}
\par\vspace{-0.08cm}\par\noindent\refjl{Coecke, B. and Stubbe,
I.}{1999a}{Operational resolutions and state transitions in a
categorical setting}{Foundations of Physics Letters}{12}{29--49}
\par\vspace{-0.08cm}\par\noindent\refjl{Coecke, B. and Stubbe, I.}{1999b}{On a duality of quantales emerging from
an operational resolution}{International Journal of Theoretical
Physics}{38}{3269--3281}
\par\vspace{-0.08cm}\par\noindent\refjl{Coecke, B. and Stubbe, I.}{2000}{State transitions as morphisms
for complete lattices}{International Journal of Theoretical
Physics}{39}{601--610}
\par\vspace{-0.08cm}\par\noindent\refpr{Cooke, R.M., and Hilgevoord, J.}{1981}{A new
approach to equivalence in quantum logic}{E. Beltrametti and B.C. van Fraassen,
(eds.)}{Current Issues in Quantum Logic}{Plenum, New York}
\par\vspace{-0.08cm}\par\noindent\refbk{Curry, H.B.}{1963}{Foundations of Mathematical
Logic}{McGraw--Hill, New York}
\par\vspace{-0.08cm}\par\noindent\refjl{D\"ahn, G.}{1968}{Attempt  of an axiomatic foundations of quantum
mechanics and more general theories. IV}{Communications in
Mathematical Physics}{9}{192--211}
\par\vspace{-0.08cm}\par\noindent\refjl{D\"ahn, G.}{1972}{The algebra generated by physical
filters}{Communications in Mathematical Physics}{28}{109--122}
\par\vspace{-0.08cm}\par\noindent\refjl{Dalla Chiara, M.L.}{1977}{Quantum logic and physical modalities}{Journal
of Philosophical Logic}{6}{391--404}
\par\vspace{-0.08cm}\par\noindent\refpr{Dalla Chiara, M.L.}{1983}{Physical implications in a Kripkean semantical
approach to physical theories}{}{Logic in the 20$^{\rm th}$ Century}{Scienta (Milan)}
\par\vspace{-0.08cm}\par\noindent\refpr{Dalla Chiara, M.L.}{1986}{Quantum logic}{D. Gabbay and F. Guenthner
(eds.)}{Handbook of Philosophical Logic}{D. Reidel, Dordrecht}
\par\vspace{-0.08cm}\par\noindent\refjl{Daniel, W.}{1982}{On a non--unitary evolution of quantum
systems}{Helvetica Physica Acta}{55}{330--338}
\par\vspace{-0.08cm}\par\noindent\refjl{Daniel, W.}{1989}{Axiomatic description of irreversible and reversible
evolution of a physical system}{Helvetica Physica
Acta}{62}{941--968}
\par\vspace{-0.08cm}\par\noindent\refjl{Derd\'erian, J.-C.}{1967}{Residuated mappings}{Pacific Journal of
Mathematics}{20}{35--43}
\par\vspace{-0.08cm}\par\noindent\refbk{Dirac, P.A.M.}{1930}{The Principles of Quantum Mechanics}{Fourth
edition. Clarendon Press, Oxford}
\par\vspace{-0.08cm}\par\noindent\refjl{Dvure\v{c}enskij, A.}{1992}{Quantum logics and completeness criteria of
inner product spaces}{International Journal of Theoretical
Physics}{31}{1899--1907}
\par\vspace{-0.08cm}\par\noindent\refbk{Dvure\v{c}enskij,
A.}{1993}{Gleason's Theorem and its Applications}{Kluwer (Dordrecht)}
\par\vspace{-0.08cm}\par\noindent\refjl{Dvure\v{c}enskij,
A.}{1995}{Tensor product of difference posets and effect algebras}{International Journal of Theoretical
Physics}{34}{1337--1348}
\par\vspace{-0.08cm}\par\noindent\refjl{Dvure\v{c}enskij, A. and Pulmannov\'a, S.}{1994a}{Tensor products of
D-posets and D-test spaces}{Reports in Mathematical
Physics}{34}{251--275}
\par\vspace{-0.08cm}\par\noindent\refjl{Dvure\v{c}enskij, A. and Pulmannov\'a, S.}{1994b}{Difference posets,
effects, and quantum measurements}{International Journal of
Theoretical Physics}{33}{819--850}
\par\vspace{-0.08cm}\par\noindent\refjl{Einstein, A., Podolsky, B. and Rosen, N.}{1935}{Can quantum--mechanical
description of reality be considered complete?}{Physcial
Review}{47}{777--780}
\par\vspace{-0.08cm}\par\noindent\refjl{Emch, G. and Piron, C.}{1962}{Note sur les  sym\'etries en th\'eorie
quantique}{Helvetica Physica Acta}{35}{542--543}
\par\vspace{-0.08cm}\par\noindent\refjl{Emch, G. and Piron, C.}{1963}{Symmetry in quantum theory}{Journal of
Mathematical Physics}{4}{469--473}
\par\vspace{-0.08cm}\par\noindent\refjl{Faure, Cl.-A. and Fr\"olicher, A.}{1993}{Morphisms of projective
geometries and of corresponding lattices}{Geometriae
Dedicata}{47}{25--40}
\par\vspace{-0.08cm}\par\noindent\refjl{Faure, Cl.-A. and Fr\"olicher, A.}{1994}{Morphisms of projective
geometries and semilinear maps}{Geometriae Dedicata}{53}{237--269}
\par\vspace{-0.08cm}\par\noindent\refjl{Faure, Cl.-A. and Fr\"olicher, A.}{1995}{Dualities for
infinite--dimensional projective geometries}{Geometriae
Dedicata}{56}{225--236}
\par\vspace{-0.08cm}\par\noindent\refjl{Faure, Cl.-A., Moore, D.J. and Piron, C.}{1995}{Deterministic evolutions
and Schr\"odinger flows}{Helvetica Physica Acta}{68}{150--157}
\par\vspace{-0.08cm}\par\noindent\refpr{Finkelstein, D.}{1968}{Matter, space and logic}{R.S. Cohen and M.W.
Wartofsky (eds.)}{Boston Studies in the Philosophy of Science
V}{D. Reidel, Dordrecht}
\par\vspace{-0.08cm}\par\noindent\refpr{Finkelstein, D.}{1972}{The physics of logic}{R.G. Colodny
(ed.)}{Paradigms {\&} Paradoxes$\,$: the Philosophical Challenge
of the Quantum Domain}{University of Pittsburgh Press, Pittsburgh}
\par\vspace{-0.08cm}\par\noindent\refbk{Foulis, D.J.}{1958}{Involution Semigroups}{Doctoral Dissertation, Tulane University}
\par\vspace{-0.08cm}\par\noindent\refjl{Foulis, D.J.}{1960}{Baer $^\ast$--semigroups}{Proceedings of the
American Mathematical Society}{11}{648--654}
\par\vspace{-0.08cm}\par\noindent\refjl{Foulis, D.J.}{1962}{A note on orthomodular lattices}{Portugaliae
Mathematica}{21}{65--72}
\par\vspace{-0.08cm}\par\noindent\refjl{Foulis, D.J.}{1998}{Mathematical metascience}{Journal of Natural
Geometry}{13}{1--50}
\par\vspace{-0.08cm}\par\noindent\refpr{Foulis, D.J.}{1999}{A half century of quantum logic ---- what have we
learned?}{D. Aerts and J. Pykacz (eds.)}{Quantum Structures and
the Nature of Reality$\,$: The Indigo Book of Einstein meets
Magritte}{Kluwer, Dordrecht}
\par\vspace{-0.08cm}\par\noindent\refpr{Foulis,
D.J.}{2000}{Representations on unigroups}{B. Coecke, D.J.
Moore and A. Wilce (eds.)}{Current Research in Operational
Quantum Logic: Algebras Categories, Languages}{Kluwer Academic
Publishers}
\par\vspace{-0.08cm}\par\noindent\refjl{Foulis, D.J., and Bennett, M.K.}{1993}{Tensor products of
orthoalgebras}{Order}{10}{271--282}
\par\vspace{-0.08cm}\par\noindent\refjl{Foulis, D.J. and Bennett, M.K.}{1994}{Effect algebras and unsharp
quantum logics}{Foundations of Physics}{24}{1331--1352}
\par\vspace{-0.08cm}\par\noindent\refjl{Foulis, D.J., Bennett, M.K. and Greechie, R.J.}{1996}{Test
groups and effect algebras}{International Journal of Theoretical
Physics}{35}{1117--1140}
\par\vspace{-0.08cm}\par\noindent\refjl{Foulis, D.J., Greechie, R.J. and Bennett, M.K.}{1998}{The transition
to unigroups}{International Journal of Theoretical
Physics}{37}{45--63}
\par\vspace{-0.08cm}\par\noindent\refjl{Foulis, D.J., Greechie, R.J. and R\"uttimann, G.T.}{1992}{Filters and
supports  in orthoalgebras}{International Journal of Theoretical
Physics}{31}{789--807}
\par\vspace{-0.08cm}\par\noindent\refjl{Foulis, D.J., Greechie, R.J. and R\"uttimann,
G.T.}{1993}{Logicoalgebraic structures II. Supports in test
spaces}{International Journal of Theoretical
Physics}{32}{1675--1690}
\par\vspace{-0.08cm}\par\noindent\refjl{Foulis, D.J., Piron, C. and Randall, C.H.}{1983}{Realism, operationalism,
and quantum mechanics}{Foundations of Physics}{13}{813--841}
\par\vspace{-0.08cm}\par\noindent\refjl{Foulis, D.J. and Randall, C.H.}{1971}{Lexicographic
orthogonality}{Journal of Combinatorial Theory}{11}{157--162}
\par\vspace{-0.08cm}\par\noindent\refjl{Foulis, D.J. and Randall, C.H.}{1972}{Operational statistics. I. Basic
concepts}{Journal of Mathematical Physics}{13}{1667--1675}
\par\vspace{-0.08cm}\par\noindent\refjl{Foulis, D.J. and Randall, C.H.}{1974}{Empirical logic and quantum
mechanics}{Synthese}{29}{81--111}
\par\vspace{-0.08cm}\par\noindent\refpr{Foulis, D.J. and Randall, C.H.}{1978}{Manuals, morphisms and quantum
mechanics}{A. Marlow (ed.)}{Mathematical Foundations of Quantum Theory}{Academic
Press, New York}
\par\vspace{-0.08cm}\par\noindent\refjl{Foulis, D.J. and Randall, C.H.}{1979}{Tensor products of quantum
logics do not exist}{Notices of the American Mathematical
Society}{26}{557}
\par\vspace{-0.08cm}\par\noindent\refpr{Foulis, D.J. and Randall,
C.H.}{1981a}{What are quantum logics and what
ought they be?}{E.G. Beltrametti and B.C. van Fraassen
(eds.)}{Current Issues in Quantum Logic}{Plenum, New York}
\par\vspace{-0.08cm}\par\noindent\refpr{Foulis, D.J. and Randall, C.H.}{1981b}{Empirical logic and tensor
products}{H. Neumann (ed.)}{Interpretations and Foundations of
Quantum Theory}{B. I. Wissenschaft, Mannheim}
\par\vspace{-0.08cm}\par\noindent\refjl{Foulis, D.J. and Randall, C.H.}{1984}{A note on
misunderstandings of Piron's axioms for quantum
mechanics}{Foundations of Physics}{14}{65--88}
\par\vspace{-0.08cm}\par\noindent\refpr{Foulis, D.J. and Wilce,
A.}{2000}{Free extensions of group actions,  induced representations,
and the foundations of physics}{B. Coecke, D.J.
Moore and A. Wilce (eds.)}{Current Research in Operational
Quantum Logic: Algebras Categories, Languages}{Kluwer Academic
Publishers}
\par\vspace{-0.08cm}\par\noindent\refpr{van Fraassen, B.C.}{1981}{Assumptions and interpretations
of quantum logic}{E.G. Beltrametti and B.C. van Fraassen
(eds.)}{Current Issues in Quantum Logic}{Plenum, New York}
\par\vspace{-0.08cm}\par\noindent\refjl{Frink, O.}{1962}{Pseudo--complements in semi--lattices}{Duke Mathematical
Journal}{29}{505--514}
\par\vspace{-0.08cm}\par\noindent\refjl{Giovannini, N.}{1981a}{Superselection variables and generalized
multipliers}{Letters in Mathematical Physics}{5}{161--168}
\par\vspace{-0.08cm}\par\noindent\refjl{Giovannini, N.}{1981b}{Classical and quantal systems of
imprimitivity}{Journal of Mathematical Physics}{22}{2389--2396}
\par\vspace{-0.08cm}\par\noindent\refjl{Giovannini, N.}{1981c}{State spaces for classical and quantal,
relativistic and nonrelativistic elementary particles}{Journal of
Mathematical Physics}{22}{2397--2403}
\par\vspace{-0.08cm}\par\noindent\refjl{Giovannini, N. and Piron, C.}{1979}{On the group--theoretical foundations
of classical and quantum physics: kinematics and state
spaces}{Helvetica Physica Acta}{52}{518--540}
\par\vspace{-0.08cm}\par\noindent\refjl{Gisin, N.}{1981}{A  simple nonlinear dissipative quantum evolution
equation}{Journal of Physics A}{14}{2259--2267}
\par\vspace{-0.08cm}\par\noindent\refjl{Gisin, N.}{1982a}{Ind\'eterminisme
quantique et dynamique non lin\'eaire dissipative}{Annales de la
Fondation Louis de Broglie}{7}{275--292}
\par\vspace{-0.08cm}\par\noindent\refjl{Gisin, N.}{1982b}{Microscopic derivation of a class of non--linear
dissipative Schr\"o\-dinger-like equations}{Physica
A}{111}{364--370}
\par\vspace{-0.08cm}\par\noindent\refpr{Gisin, N.}{1983a}{Mod\`eles d'un processus de mesure}{C. Gruber, C.
Piron, T.M. T\^am and R. Weill (eds.)}{Les fondements de la
m\'ecanique quantique}{AVCP, Lausanne}
\par\vspace{-0.08cm}\par\noindent\refjl{Gisin, N.}{1983b}{Irreversible quantum
dynamics and the Hilbert space structure of quantum
kinematics}{Journal of Mathematical Physics}{24}{1779--1782}
\par\vspace{-0.08cm}\par\noindent\refjl{Gisin, N. and Piron, C.}{1981}{Collapse of the wave packet without
mixture}{Letters in Mathematical Physics}{5}{379--385}
\par\vspace{-0.08cm}\par\noindent\refjl{Giuntini, R. and Greuling, H.}{1989}{Toward a formal
language for unsharp properties}{Foundations of
Physics}{19}{931--945}
\par\vspace{-0.08cm}\par\noindent\refjl{Gleason, A.M.}{1957}{Measures on the closed subspaces of a
Hilbert space}{Journal of Mathematics and Mechanics}{6}{885--893}
\par\vspace{-0.08cm}\par\noindent\refjl{Goldblatt, R.I.}{1974}{Semantic analysis of
orthologic}{Journal of Philosophical Logic}{3}{19--35}
\par\vspace{-0.08cm}\par\noindent\refjl{Goldblatt, R.I.}{1975}{The Stone space of an
ortholattice}{Bulletin of the London Mathematical
Society}{7}{45--48}
\par\vspace{-0.08cm}\par\noindent\refbk{Golfin,
A.S.Jr.}{1987}{Representations and Products of Lattices}{Doctoral
Dissertation, University of Massachusetts}
\par\vspace{-0.08cm}\par\noindent\refjl{Greechie, R.J., Foulis, D.J. and Pulmannov\'a, S.}{1995}{The
center of an effect algebra}{Order}{12}{91--106}
\par\vspace{-0.08cm}\par\noindent\refbk{Gross, H.}{1979}{Quadratic Forms in Infinite Dimensional Vector
Spaces}{Birkhauser, Boston}
\par\vspace{-0.08cm}\par\noindent\refjl{Gross, H.}{1990}{Hilbert lattices: new results and
unsolved problems}{Foundations of  Physics}{20}{529--559}
\par\vspace{-0.08cm}\par\noindent\refjl{Gudder, S.P.}{1965}{Spectral methods for generalized
probability theory}{Transactions of the American Mathematical
Society}{119}{428--442}
\par\vspace{-0.08cm}\par\noindent\refpr{Gudder, S.P.}{1977}{Four approaches to axiomatic  quantum mechanics}{W.C.
Price and S.S. Chissick (eds.)}{The Uncertainty Principle and
Foundations of Quantum Mechanics}{John Wiley \& Sons, London}
\par\vspace{-0.08cm}\par\noindent\refpr{Gudder, S.P.}{1979}{A survey of axiomatic quantum mechanics}{C.A.
Hooker (ed.)}{The Logico--Algebraic Approach to Quantum Mechanics.
II.}{D. Reidel, Dordrecht}
\par\vspace{-0.08cm}\par\noindent\refpr{Gudder, S.P.}{1981}{Comparison of the quantum logic, convexity, and
algebraic approaches to quantum mechanics}{H. Neumann
(ed.)}{Interpretations and Foundations of Quantum Theory}{B. I.
Wissenschaft, Mannheim}
\par\vspace{-0.08cm}\par\noindent\refbk{Gudder, S.P.}{1985}{Quantum Probability}{Academic
Press, San Diego}
\par\vspace{-0.08cm}\par\noindent\refjl{Gudder, S.P.}{1997}{Effect test spaces}{International Journal
of Theoretical Physics}{36}{2681--2705}
\par\vspace{-0.08cm}\par\noindent\refpr{Gudder, S.}{2000}{Quantum
languages}{B. Coecke, D.J.
Moore and A. Wilce (eds.)}{Current Research in Operational
Quantum Logic: Algebras Categories, Languages}{Kluwer Academic
Publishers}
\par\vspace{-0.08cm}\par\noindent\refjl{Gudder, S.P. and Pulmannov\'a, S.}{1987}{Transition amplitude
spaces}{Journal of Mathematical Physics}{28}{376--385}
\par\vspace{-0.08cm}\par\noindent\refjl{Haag, R. and Kastler, D.}{1964}{An algebraic approach to quantum field
theory}{Journal of Mathematical Physics}{5}{848----861}
\par\vspace{-0.08cm}\par\noindent\refbk{Habil, E.}{1993}{Orthoalgebras and Noncommutative Measure
Theory}{Doctoral Dissertation, Kansas State University}
\par\vspace{-0.08cm}\par\noindent\refjl{Hamhalter, J.}{1993}{Pure Jauch--Piron states on von Neumann
algebras}{Annales de l'Institut Henri Poincar\'e A}{58}{173--187}
\par\vspace{-0.08cm}\par\noindent\refjl{Hamhalter, J.}{1995}{Extension properties of states on operator
algebras}{International Journal of Theoretical
Physics}{34}{1431--1437}
\par\vspace{-0.08cm}\par\noindent\refpr{Hardegree, G.M. and Frazer, P.J.}{1981}{Charting the labyrinth of
quantum logics: a progress report}{E.G. Beltrametti and B.C. van
Fraassen (eds.)}{Current Issues in Quantum Logic}{Plenum, New
York}
\par\vspace{-0.08cm}\par\noindent\refjl{Harding, J.}{1996}{Decompositions in quantum logic}{Transactions of the American
Mathematical Society}{348}{1839--1862}
\par\vspace{-0.08cm}\par\noindent\refjl{Harding, J.}{1998}{Regularity in quantum logic}{International Journal of
Theoretical Physics}{37}{1173--1212}
\par\vspace{-0.08cm}\par\noindent\refjl{Herman, L., Marsden, E.L. and Piziak, R.}{1975}{Implication
connectives in orthomodular lattices}{Notre Dame Journal of Formal
Logic}{16}{305--328}
\par\vspace{-0.08cm}\par\noindent\refjl{Holland, S.S.}{1995}{Orthomodularity in infinite dimensions: a
theorem of M. Sol\`er}{Bulletin of the American Mathematical
Society}{32}{205--234}
\par\vspace{-0.08cm}\par\noindent\refbk{Jauch, J.M.}{1968}{Foundations of Quantum
Mechanics}{Addison--Wesley, Reading}
\par\vspace{-0.08cm}\par\noindent\refjl{Jauch, J.M. and Piron, C.}{1967}{Generalized localizability}{Helvetica
Physica Acta}{40}{559--570}
\par\vspace{-0.08cm}\par\noindent\refjl{Jauch, J.M. and Piron, C.}{1969}{On the structure of quantal proposition
systems}{Helvetica Physica Acta}{42}{842--848}
\par\vspace{-0.08cm}\par\noindent\refbk{Joyal, A. and Tierney, M.}{1984}{An extension of the Galois theory of
Grothendieck}{American Mathematical Society, Providence}
\par\vspace{-0.08cm}\par\noindent\refjl{Kalmbach, G.}{1974}{Orthomodular logic}{Zeitschrift f\"ur
Mathematische Logik und Grundlagen der Mathematik}{20}{395--406}
\par\vspace{-0.08cm}\par\noindent\refbk{Kalmbach, G.}{1983}{Orthomodular Lattices}{Academic
Press, London}
\par\vspace{-0.08cm}\par\noindent\refjl{Keller, H.A.}{1980}{Ein nicht--klassischer Hilbertscher
Raum}{Mathematische Zeietschrift}{172}{41--49}
\par\vspace{-0.08cm}\par\noindent\refbk{Kelly, G.M.}{1982}{Basic Concepts of Enriched Category
Theory}{Cambridge University Press, Cambridge}
\par\vspace{-0.08cm}\par\noindent\refjl{Kl\"{a}y, M., Randall, C.H. and Foulis, D.J.}{1987}{Tensor products
and probability weights}{International Journal of Theoretical
Physics}{26}{199--219}
\par\vspace{-0.08cm}\par\noindent\refpr{Kochen, S.}{1996}{Construction
of quantum mechanics via commutatitive operations}{R. Clifton
(ed.)} {Perspectives on Quantum Reality}{Kluwer Academic
Publishers}
\par\vspace{-0.08cm}\par\noindent\refjl{Kochen, S. and Specker,
E.P.}{1967}{The problem of hidden variables in quantum
mechanics}{Journal of Mathematics and Mechanics}{17}{59--87}
\par\vspace{-0.08cm}\par\noindent\refjl{K\"ohler, P.}{1981}{Brouwerian semilattices}{Transactions of the
American Mathematical Society}{268}{103--126}
\par\vspace{-0.08cm}\par\noindent\refjl{K\^opka, F.}{1992}{D-posets of fuzzy sets}{Tatra Mountains
Mathematical Publications}{1}{83--87}
\par\vspace{-0.08cm}\par\noindent\refjl{Kruml, D}{2000}{Spatial quantales}{Applied Categorical
Structures}{}{To appear}
\par\vspace{-0.08cm}\par\noindent\refbk{Loomis, L.}{1955}{The lattice--theoretic background of
the Dimension Theory of Operator Algebras}{American Mathematical
Society, Providence}
\par\vspace{-0.08cm}\par\noindent\refbk{Ludwig, G.}{1954}{Die Grundlagen der
Quantenmechanik}{Springer--Verlag, Berlin, Translation by C.~A. Hein ``Foundations of
Quantum Mechanics'' Springer--Verlag, Berlin, 1983}
\par\vspace{-0.08cm}\par\noindent\refjl{Ludwig, G.}{1955}{Zur Deutung der Beobachung in der
Quantenmechanik}{Physik\-alisch Bl\"atter}{11}{489--494}
\par\vspace{-0.08cm}\par\noindent\refjl{Ludwig, G.}{1964}{Versuch einer axiomatischen Grundlegung der
Quantenmechanik und allgemeiner physikalischer
Theorien}{Zeitschrift f\"ur Physik}{181}{233--260}
\par\vspace{-0.08cm}\par\noindent\refjl{Ludwig, G.}{1967}{Attempt of an axiomatic foundations of quantum
mechanics and more general theories. II}{Communications in
Mathematical Physics}{4}{331--348}
\par\vspace{-0.08cm}\par\noindent\refjl{Ludwig, G.}{1968}{Attempt of an axiomatic foundations of quantum
mechanics and more general theories. III}{Communications in
Mathematical Physics}{9}{1--12}
\par\vspace{-0.08cm}\par\noindent\refjl{Ludwig, G.}{1972}{An improved  formulation of some theorems and axioms
in the axiomatic foundation of the Hilbert space structure of
quantum mechanics}{Communications in Mathematical
Physics}{26}{78--86}
\par\vspace{-0.08cm}\par\noindent\refbk{Ludwig, G.}{1985}{An Axiomatic Basis of Quantum Mechanics. 1.
Derivation of Hilbert Space}{Springer--Verlag, Berlin}
\par\vspace{-0.08cm}\par\noindent\refbk{Ludwig, G.}{1987}{An Axiomatic Foundation of Quantum
Mechanics. 2. Quantum Mechanics and Macrosystems}{Springer--Verlag, Berlin}
\par\vspace{-0.08cm}\par\noindent\refpr{Ludwig, G. and Neumann, H.}{1981}{Connections  between different
approaches to the foundations of quantum mechanics}{H. Neumann
(ed.)}{Interpretations and Foundations of Quantum Theory}{B. I.
Wissenschaft, Mannheim}
\par\vspace{-0.08cm}\par\noindent\refjl{Mackey, G.W.}{1957}{Quantum mechanics and Hilbert space}{American
Mathematical Monthly}{64:2}{45--57}
\par\vspace{-0.08cm}\par\noindent\refbk{Mackey, G.W.}{1963}{The Mathematical Foundations of Quantum
Mechanics}{W.~A. Benjamin, New York}
\par\vspace{-0.08cm}\par\noindent\refjl{Maeda, S.}{1955}{Dimension functions on certain general
lattices}{Journal of Science of Hiroshima University
A}{19}{211--237}
\par\vspace{-0.08cm}\par\noindent\refjl{Malinowski, J.}{1990}{The deduction theorem for quantum logic---some
negative results}{Journal of Symbolic Logic}{55}{615--625}
\par\vspace{-0.08cm}\par\noindent\refjl{Malinowski, J.}{1992}{Strong
versus weak quantum consequence operations}{Studia Logica}{51}{113--123}
\par\vspace{-0.08cm}\par\noindent\refjl{Mielnik, B.}{1968}{Geometry of quantum states}{Communications in
Mathematical Physics}{9}{55--80}
\par\vspace{-0.08cm}\par\noindent\refjl{Mielnik, B.}{1969}{Theory of filters}{Communications in Mathematical
Physics}{15}{1--46}
\par\vspace{-0.08cm}\par\noindent\refjl{Moore, D.J.}{1993}{Quantum logic requires weak
modularity}{Helvetica Physica Acta}{66}{471--476}
\par\vspace{-0.08cm}\par\noindent\refjl{Moore, D.J.}{1995}{Categories of representations of physical
systems}{Helvetica Physica Acta}{68}{658--678}
\par\vspace{-0.08cm}\par\noindent\refjl{Moore, D.J.}{1997}{Closure categories}{International Journal of
Theoretical Physics}{36}{2707--2723}
\par\vspace{-0.08cm}\par\noindent\refjl{Moore, D.J.}{1999}{On state spaces and property lattices}{Studies in the
History and Philosophy of Modern Physics}{30}{61--83}
\par\vspace{-0.08cm}\par\noindent\refjl{Mulvey, C.J.}{1986}{$\&$}{Supplemento ai Rendiconti del Circolo Mathematico di
Palermo}{12}{99--104}
\par\vspace{-0.08cm}\par\noindent\refjl{Nakamura, M.}{1957}{The permutability in a certain orthocomplemented
lattice}{Kodai Mathematical Seminar Reports}{9}{158--160}
\par\vspace{-0.08cm}\par\noindent\refjl{Navara, M.}{1992}{Independence of automorphism group, center, and state
space of quantum logics}{International Journal of Theoretical
Physics}{31}{925--935}
\par\vspace{-0.08cm}\par\noindent\refjl{Nemitz, W.C.}{1965}{Implicative semi--lattices}{Transactions of the
American Mathematical Society}{117}{128--142}
\par\vspace{-0.08cm}\par\noindent\refbk{von Neumann, J.}{1932}{Mathematische Grundlagen der
Quantenmechanik}{Springer--Verlag, Berlin. Translation by R.
Beyer ``Mathematical Foundations of Quantum Mechanics'' Princeton
University Press, Princeton, 1955}
\par\vspace{-0.08cm}\par\noindent\refjl{Nishimura, H.}{1993}{Empirical set
theory}{International Journal of Theoretical Physics}{32}{1293--1321}
\par\vspace{-0.08cm}\par\noindent\refjl{Nishimura, H.}{1995}{Manuals in orthogonal
categories}{International Journal of Theoretical Physics}{34}{211--228}
\par\vspace{-0.08cm}\par\noindent\refpr{Paseka, J.}{1997}{Simple
quantales}{}{Proceedings of the Eighth Prague Topology Symposium}{Prague}
\par\vspace{-0.08cm}\par\noindent\refjl{Paseka, J. and Kruml, D.}{2000}{Embeddings of quantales
into simple quantales}{Journal of Pure and Applied Algebra}{}{To
appear}
\par\vspace{-0.08cm}\par\noindent\refpr{Paseka, J. and Rosick\'y,
J.}{2000}{Quantales}{B. Coecke, D.J.
Moore and A. Wilce (eds.)}{Current Research in Operational
Quantum Logic: Algebras Categories, Languages}{Kluwer Academic
Publishers}
\par\vspace{-0.08cm}\par\noindent\refjl{Piron, C.}{1964}{Axiomatique quantique}{Helvetica Physica
Acta}{37}{439--468}
\par\vspace{-0.08cm}\par\noindent\refjl{Piron, C.}{1965}{Sur  la quantification du syst\`eme de deux
particules}{Helvetica Physica Acta}{38}{104--108}
\par\vspace{-0.08cm}\par\noindent\refjl{Piron, C.}{1969}{Les r\`egles de supers\'election continues}{Helvetica
Physica Acta}{42}{330--338}
\par\vspace{-0.08cm}\par\noindent\refpr{Piron, C.}{1971}{Observables in general  quantum theory}{B. D'Espagnat
(ed.)}{Foundations of Quantum Mechanics}{Academic Press, New
York}
\par\vspace{-0.08cm}\par\noindent\refjl{Piron, C.}{1972}{Survey of general quantum physics}{Foundations of
Physics}{2}{287--314}
\par\vspace{-0.08cm}\par\noindent\refbk{Piron, C.}{1976}{Foundations of Quantum
Mechanics}{W.A. Benjamin, Inc., Reading}
\par\vspace{-0.08cm}\par\noindent\refbk{Piron, C.}{1990}{M\'ecanique quantique bases et applications}{Presses
polytechniques et universitaires romandes, Lausanne, Second
edition, 1998}
\par\vspace{-0.08cm}\par\noindent\refjl{Pitts, A.M.}{1988}{Applications of sub--lattice enriched category theory
to sheaf theory}{Proceedings of the London Mathematical
Society}{57}{433--480}
\par\vspace{-0.08cm}\par\noindent\refjl{Pool, J.C.T.}{1968a}{Baer$^\ast$--semigroups and the logic of
quantum mechanics}{Communications in Mathematical
Physics}{9}{118--141}
\par\vspace{-0.08cm}\par\noindent\refjl{Pool, J.C.T.}{1968b}{Semimodularity and the logic of quantum
mechanics}{Communications in Mathematical Physics}{9}{212--228}
\par\vspace{-0.08cm}\par\noindent\refjl{Prestel, A.}{1995}{On Sol\`er's characterization of Hilbert
spaces}{Manususcripta Mathematica}{86}{225--238}
\par\vspace{-0.08cm}\par\noindent\refpr{Pt\'ak, P.}{2000}{Observables
in the logico-algebraic approach}{B. Coecke, D.J.
Moore and A. Wilce (eds.)}{Current Research in Operational
Quantum Logic: Algebras Categories, Languages}{Kluwer Academic
Publishers}
\par\vspace{-0.08cm}\par\noindent\refbk{Pt\'ak, P. and Pulmannov\'a, S.}{1991}{Orthomodular Structures as Quantum
Logics}{Kluwer, Dordrecht}
\par\vspace{-0.08cm}\par\noindent\refjl{Pulmannov\'a, S.}{1986a}{Transition probability spaces}{Journal of
Mathematical Physics}{27}{1791--1795}
\par\vspace{-0.08cm}\par\noindent\refjl{Pulmannov\'a, S.}{1986b}{Functional properties of transition
amplitude spaces}{Reports in  Mathematical Physics}{24}{81--86}
\par\vspace{-0.08cm}\par\noindent\refjl{Pulmannov\'a, S. and Gudder, S.P.}{1987}{Geometric properties of transition
amplitude spaces}{Journal of Mathematical Physics}{28}{2393--2399}
\par\vspace{-0.08cm}\par\noindent\refjl{Pulmannov\'a, S. and Wilce, A.}{1995}{Representations of D--posets}{International
Journal of Theoretical Physics}{34}{1689--1696}
\par\vspace{-0.08cm}\par\noindent\refpr{Putnam, H.}{1968}{Is logic empirical?}{R.S.
Cohen and M.W. Wartofsky (eds.)}{Boston Studies in the Philosophy
of Science V}{D. Reidel, Dordrecht}
\par\vspace{-0.08cm}\par\noindent\refpr{Putnam, H.}{1976}{How to think quantum--logically}{P. Suppes (ed.)}{Logic
and Probability in Quantum Mechanics}{D. Reidel, Dordrecht}
\par\vspace{-0.08cm}\par\noindent\refbk{Randall, C.H.}{1966}{A Mathematical Foundation for
Empirical Science}{Doctoral Dissertation, Rensselaer Polytechnic
Institute}
\par\vspace{-0.08cm}\par\noindent\refjl{Randall, C.H. and Foulis, D.J.}{1970}{An approach to empirical
logic}{American Mathematical Monthly}{77}{363--374}
\par\vspace{-0.08cm}\par\noindent\refjl{Randall, C.H. and Foulis, D.J.}{1973}{Operational statistics. II.
Manuals of  operations and their logics}{Journal of Mathematical
Physics}{14}{1472--1480}
\par\vspace{-0.08cm}\par\noindent\refpr{Randall, C.H. and Foulis, D.J.}{1978}{The operational approach to
quantum mechanics}{C.A. Hooker (ed.)}{Physical Theory as
Logico-Operational Structure}{D. Reidel, Dordrecht}
\par\vspace{-0.08cm}\par\noindent\refpr{Randall, C.H. and Foulis, D.J.}{1981}{Operational statistics and
tensor products}{H. Neumann (ed.)}{Interpretations and Foundations
of Quantum Theory}{B.~I. Wissenschaft, Mannheim}
\par\vspace{-0.08cm}\par\noindent\refpr{Randall, C.H. and Foulis, D.J.}{1983a}{A mathematical language for
quantum physics}{C. Gruber, C. Piron, T.M. T\^am and R. Weill
(eds.)}{Les fondements de la m\'ecanique quantique}{AVCP, Lausanne}
\par\vspace{-0.08cm}\par\noindent\refjl{Randall, C.H. and Foulis, D.J.}{1983b}{Properties and operational
propositions in quantum mechanics}{Foundations of
Physics}{13}{843--857}
\par\vspace{-0.08cm}\par\noindent\refpr{Resende, P.}{1999}{Modular specification of concurrent systems with
observational logic}{J.L. Fiadeiro (ed.)}{Recent Developments in
Algebraic Development Techniques}{Springer--Verlag, Berlin}
\par\vspace{-0.08cm}\par\noindent\refjl{Resende, P.}{2000}{Quantales, finite observations and strong
bisimulation}{Theoretical Computer Science}{}{To appear}
\par\vspace{-0.08cm}\par\noindent\refpr{Resende, P.}{2000}{Quantales
and observational semantics}{B. Coecke, D.J.
Moore and A. Wilce (eds.)}{Current Research in Operational
Quantum Logic: Algebras Categories, Languages}{Kluwer Academic
Publishers}
\par\vspace{-0.08cm}\par\noindent\refjl{Rosenthal, K.I.}{1991}{Free quantaloids}{Journal of Pure and Applied
Algebra}{77}{67--82}
\par\vspace{-0.08cm}\par\noindent\refjl{Rosick\'y, J.}{1989}{Multiplicative lattices and C$^\ast$--algebras}{Cahiers
de Topologie et G\'eom\'etrie Diff\'erentielle
Cat\'egoriques}{30}{95--110}
\par\vspace{-0.08cm}\par\noindent\refjl{Rosick\'y, J.}{1995}{Characterizing spatial
quantales}{Algebra Universalis}{34}{175--178}
\par\vspace{-0.08cm}\par\noindent\refjl{Sasaki, U.}{1955}{Lattices of projections in AW$^\ast$--algebras}{Journal
of Science of Hiroshima Univerity A}{19}{1--30}
\par\vspace{-0.08cm}\par\noindent\refbk{Schroeck, F.E.}{1996}{Quantum
Mechanics on Phase Space}{Kluwer Academic (Dordrecht)}
\par\vspace{-0.08cm}\par\noindent\refjl{Segal, I.}{1947}{Postulates for
General Quantum Mechanics}{Annals of Mathematics}{4}{930--948}
\par\vspace{-0.08cm}\par\noindent\refjl{Sol\`er, M.P.}{1995}{Characterization of Hilbert spaces by orthomodular
spaces}{Communcations in Algebra}{23}{219--243}
\par\vspace{-0.08cm}\par\noindent\refjl{Stolz, P.}{1969}{Attempt of an axiomatic foundation of quantum
mechanics and more general theories. V}{Communications in
Mathematical Physics}{11}{303--313}
\par\vspace{-0.08cm}\par\noindent\refjl{Stolz, P.}{1971}{Attempt of an axiomatic foundation of quantum mechanics
and more general theories. VI}{Communications in Mathematical
Physics}{23}{117--126}
\par\vspace{-0.08cm}\par\noindent\refpr{Valckenborgh,
F.}{2000}{Operational axiomatics and compound systems}{B. Coecke, D.J.
Moore and A. Wilce (eds.)}{Current Research in Operational
Quantum Logic: Algebras Categories, Languages}{Kluwer Academic
Publishers}
\par\vspace{-0.08cm}\par\noindent\refbk{Varadarajan, V.S.}{1968}{The
Geometry of Quantum Theory, Volumes I--II}{Van Nostrand, New York}
\par\vspace{-0.08cm}\par\noindent\refbk{Vickers, S.}{1989}{Topology
via Logic}{Cambridge University
Press, Cambridge}
\par\vspace{-0.08cm}\par\noindent\refjl{Ward, M.}{1937}{Residuations in structures over which a multiplication is
defined}{Duke Mathematical Journal}{3}{627--636}
\par\vspace{-0.08cm}\par\noindent\refjl{Ward, M.}{1938}{Structure residuation}{Annals of
Mathematics}{39}{558--568}
\par\vspace{-0.08cm}\par\noindent\refjl{Ward, M. and Dilworth, R.P.}{1939a}{Residuated lattices}{Transactions of
the American Mathematical Society}{45}{335--354}
\par\vspace{-0.08cm}\par\noindent\refjl{Ward, M. and Dilworth, R.P.}{1939b}{Non--commutative residuated
lattices}{Transactions of the American Mathematical
Society}{46}{426--444}
\par\vspace{-0.08cm}\par\noindent\refjl{Wilce, A.}{1990}{Tensor products of frame manuals}{International Journal of
Theoretical Physics}{29}{805--814}
\par\vspace{-0.08cm}\par\noindent\refjl{Wilce, A.}{1992}{Tensor products in generalized measure theory}{International
Journal of Theoretical Physics}{31}{1915--1928}
\par\vspace{-0.08cm}\par\noindent\refjl{Wilce, A.}{1995}{Partial Abelian semigroups}{International Journal of Theoretical
Physics}{34}{1807--1812}
\par\vspace{-0.08cm}\par\noindent\refjl{Wilce, A.}{1997}{Pull-backs and product tests}{Helvetica Physica Acta}{70}{803--812}
\par\vspace{-0.08cm}\par\noindent\refjl{Wilce, A.}{1998}{Perspectivity and congruence in partial Abelian
semigroups}{Mathematica Slovaca}{48}{117--135}
\par\vspace{-0.08cm}\par\noindent\refjl{Wilce, A.}{2000}{On generalized Sasaki
projections}{International Journal of Theoretical
Physics}{39}{To appear}
\par\vspace{-0.08cm}\par\noindent\refpr{Wilce, A.}{2000}{Test spaces
and orthoalgebras}{B. Coecke, D.J.
Moore and A. Wilce (eds.)}{Current Research in Operational
Quantum Logic: Algebras Categories, Languages}{Kluwer Academic
Publishers}
\par\vspace{-0.08cm}\par\noindent\refjl{Yeadon, F.J.}{1983}{Measures on
projections in $W^{*}$-algebras of type ${\rm II}_{1}$}{Bulletin
of the London Mathematical Society}{15}{139--145}
\end{itemize}

\end{document}